\documentclass[aps,prl,twocolumn,superscriptaddress,showpacs]{revtex4-2}
\usepackage{amssymb}
\usepackage{graphicx}
\usepackage{amsmath}
\usepackage{mathtools}
\usepackage[export]{adjustbox}
\usepackage{epsfig}
\usepackage{times}
\usepackage{xcolor}
\usepackage{subfigure}
\usepackage{setspace}
\usepackage{bm}% bold math
\usepackage{calc}
\usepackage{natbib}
\usepackage[normalem]{ulem}
\usepackage{cancel}
\usepackage{braket}

\usepackage[pdftex]{hyperref}
\hypersetup{colorlinks = true, urlcolor = blue, linkcolor = blue, citecolor = blue}

\begin{document}
%-----------------------------------------------------------------------------------------

\definecolor{ao}{rgb}{0.0, 0.5, 0.0}
\newcommand{\acc}[1]{\textcolor{red}{\textit{[AC: #1]}}}
\newcommand{\ac}[1]{\textcolor{magenta}{\textit{ #1}}}
\newcommand{\acd}[1]{\textcolor{ao}{\textit{ #1}}}
\newcommand{\fm}[1]{\textcolor{black}{#1}}
\newcommand {\ba} {\ensuremath{b^\dagger}}
\newcommand {\Ma} {\ensuremath{M^\dagger}}
\newcommand {\psia} {\ensuremath{\psi^\dagger}}
\newcommand {\psita} {\ensuremath{\tilde{\psi}^\dagger}}
\newcommand{\lp} {\ensuremath{{\lambda '}}}
\newcommand{\A} {\ensuremath{{\bf A}}}
\newcommand{\Q} {\ensuremath{{\bf Q}}}
\newcommand{\kk} {\ensuremath{{\bf k}}}
\newcommand{\qq} {\ensuremath{{\bf q}}}
\newcommand{\kp} {\ensuremath{{\bf k'}}}
\newcommand{\rr} {\ensuremath{{\bf r}}}
\newcommand{\rp} {\ensuremath{{\bf r'}}}
\newcommand {\ep} {\ensuremath{\epsilon}}
\newcommand{\nbr} {\ensuremath{\langle ij \rangle}}
\newcommand {\no} {\nonumber}
\newcommand{\up} {\ensuremath{\uparrow}}
\newcommand{\dn} {\ensuremath{\downarrow}}
\newcommand{\rcol} {\textcolor{red}}

\newcommand{\tblk}{\makebox[0pt] \boxed{~T~}}
\newcommand{\fpc}[1]{\textcolor{blue}{\textit{[FP: #1]}}}
\newcommand{\fpm}[1]{\textcolor{blue}{#1}}
\newcommand{\fps}[1]{\textcolor{blue}{\sout{#1}}}

\newcommand{\yueqing}[1]{\textcolor{orange}{\bf [YC: #1]}}
\newcommand{\fkc}[1]{\textcolor{cyan}{\bf [FK: #1]}}
\newcommand{\fk}[1]{\textcolor{cyan}{#1}}
\newcommand{\aw}[1]{\textcolor{blue}{\bf [AW: #1]}}
%-----------------------------------------------------------------------------------------

\begin{abstract}
In single sheets of graphene, vacancy-induced states have been shown to host an effective spin-1/2 hole that can be Kondo-screened at low temperatures. Here, we show how these vacancy-induced impurity states survive in twisted bilayer graphene (TBG), which thus provides a tunable system to probe the critical destruction of the Kondo effect in pseudogap hosts. Ab-initio calculations and atomic-scale modeling are used to determine the nature of the vacancy states in the vicinity of the magic angle in TBG, demonstrating that the vacancy can be treated as a quantum impurity. Utilizing this insight, we construct an Anderson impurity model with a TBG host that we solve using the numerical renormalization group combined with the kernel polynomial method. We determine the phase diagram of the model and show how there is a strict dichotomy between vacancies in the AA/BB versus AB/BA tunneling regions. 
In AB/BA vacancies, the Kondo temperature at the magic angle develops a broad distribution with a tail to vanishing temperatures due to multifractal wavefunctions at the magic angle.
We argue that scanning tunneling microscopy in the vicinity of the vacancy can act as a probe of both the critical single-particle states and the underlying many-body ground state in magic-angle TBG.
 \end{abstract}
%-----------------------------------------------------------------------------------------
\title{Vacancy-induced tunable Kondo effect in twisted bilayer graphene}
\author{Yueqing Chang}
\email{yueqing.chang@rutgers.edu}
 \affiliation{Department of Physics and Astronomy,   
Rutgers University, Piscataway, NJ 08854, USA}
  \affiliation{Center for Materials Theory,  
Rutgers University, Piscataway, NJ 08854, USA}
\author{Jinjing Yi}
 \affiliation{Department of Physics and Astronomy,   
Rutgers University, Piscataway, NJ 08854, USA}
  \affiliation{Center for Materials Theory,  
Rutgers University, Piscataway, NJ 08854, USA}
\author{Ang-Kun Wu}
  \affiliation{Department of Physics and Astronomy,   
Rutgers University, Piscataway, NJ 08854, USA}
  \affiliation{Center for Materials Theory,  
Rutgers University, Piscataway, NJ 08854, USA}
\author{Fabian B.~Kugler}
\affiliation{Center for Computational Quantum Physics, Flatiron Institute, 162 5th Avenue, New York, NY 10010, USA}
\affiliation{Department of Physics and Astronomy,   
Rutgers University, Piscataway, NJ 08854, USA}
  \affiliation{Center for Materials Theory,  
Rutgers University, Piscataway, NJ 08854, USA}
\author{\\Eva Y. Andrei}
 \affiliation{Department of Physics and Astronomy, 
Rutgers University, Piscataway, NJ 08854, USA}
\author{David Vanderbilt}
\affiliation{Department of Physics and Astronomy,   
Rutgers University, Piscataway, NJ 08854, USA}
  \affiliation{Center for Materials Theory,  
Rutgers University, Piscataway, NJ 08854, USA}
\author{Gabriel Kotliar}
\affiliation{Department of Physics and Astronomy,   
Rutgers University, Piscataway, NJ 08854, USA}
  \affiliation{Center for Materials Theory,  
Rutgers University, Piscataway, NJ 08854, USA}
\affiliation{Condensed Matter Physics and Materials Science Department,\looseness=-1\,  
Brookhaven National Laboratory, Upton, NY 11973, USA}
\author{J.~H.~Pixley}
\email{jed.pixley@physics.rutgers.edu}
\affiliation{Department of Physics and Astronomy,   
Rutgers University, Piscataway, NJ 08854, USA}
  \affiliation{Center for Materials Theory,  
Rutgers University, Piscataway, NJ 08854, USA}
\affiliation{Center for Computational Quantum Physics, Flatiron Institute, 162 5th Avenue, New York, NY 10010, USA}

\date{\today}

\maketitle

Twisted van der Waals heterostructures have taken the condensed matter community by storm~\cite{andrei_graphene_2020, balents_superconductivity_2020, andrei_marvels_2021, nuckolls_microscopic_2024}.
Since the first experimental evidence of the band reconstruction and emergence of a flat band in twisted bilayer graphene (TBG) at twist angle $\sim\!1^{\circ}$~\cite{li_observation_2010}, a wide range of experimental and technical breakthroughs~\cite{luican_single-layer_2011, yan_angle-dependent_2012, kim_van_2016, cao_superlattice-induced_2016} have paved the way for the discovery and reproducible observations of correlated insulating states~\cite{cao_correlated_2018} and superconductivity~\cite{cao_unconventional_2018} in magic-angle TBG, highlighting the vast potential and intriguing properties of these moir\'e materials~\cite{
efimkin_helical_2018,
xu_topological_2018,
kang_symmetry_2018,
koshino_maximally_2018,
padhi_doped_2018, guinea_electrostatic_2018,
tarnopolsky_origin_2019, liu_pseudo_2019, kang_strong_2019, ming_nature_2020,
bultinck_ground_2020, vafek_renormalization_2020, 
bernevig_twisted_2021_i, song_twisted_2021_ii,
bernevig_twisted_III_2021, lian_twisted_2021, bernevig_twisted_V_2021, vafek_lattice_2021, song_magic-angle_2022, hu_kondo_2023, chou_kondo_2023, calugaru_tbg_2023, lau_topological_2023, zhou_kondo_2023, chou_scaling_2023, hu_symmetric_2023}. 
These ideas have now been extended to twisted bilayer transition metal dichalcogenides (TMD)~\cite{wu_hubbard_2018, regan_mott_2020, zhang_flat_2020, wang_correlated_2020, ghiotto_quantum_2021}, quantum magnets~\cite{tong_skyrmions_2018, hejazi_noncollinear_2020}, high-temperature 
superconductors~\cite{can_high-temperature_2021, zhu_presence_2021, volkov_magic_2023, volkov_current-_2023, zhao_time_2023}, and bosonic superfluids in optical lattices~\cite{gonzalez-tudela_cold_2019, fu2020magic, luo_spin-twisted_2021, meng_atomic_2023}.
To unravel the nature of the underlying many-body states in moir\'e materials, it is essential to explore new ways to extract electron correlations while using current experimental capabilities.

One potentially fruitful direction is to probe the nature of impurity states, accurately measurable with scanning tunneling microscopy (STM), to gain insights into the many-body ground state that the impurity states are coupled to.
For instance, tunneling spectra of impurity-induced resonances in superconductors reveal signatures of the pairing symmetry~\cite{sukhachov_tunneling_2023}.
For single sheets of graphene, creating impurity states strongly coupled to the itinerant electrons was a challenge until it was realized that vacancy-induced bound states act like a spin-1/2 hole, originating from the vacancy's nearest-neighbor $\sigma$ states which couple to the $\pi$-band due to the local curvature near the vacancy site~\cite{haase_magnetic_2011, palacios_critical_2012, nanda_electronic_2012, mao_realization_2016, jiang_inducing_2018, may_modeling_2018}. 
This represents a clear-cut realization of the pseudogap Anderson impurity model (AIM), which features
a quantum critical point at non-zero Kondo coupling~\cite{PhysRevLett.64.1835, Bulla1997, 
PhysRevB.57.14254, ingersent_critical_2002, PhysRevB.70.214427,
vojta_upper_2004, fritz_universal_2006,
glossop_critical_2011, pixley_quantum_2013, fritz2013physics, pixley_entanglement_2015, wagner_long-range_2018, cai_critical_2020}.
Yet, experimentally observing this quantum critical point has remained out of reach due to the lack of tunability of vacancy states, despite the observation of Kondo screening in graphene hosts~\cite{chen2011tunable,jiang_inducing_2018}. 

In this work, 
we study vacancy-induced impurity states in TBG away from and at the magic angle with atomic-scale and effective lattice models. 
Using ab-initio calculations, we show that a vacancy~\cite{wehling2014dirac} induces an effective spin-1/2 hole on the atomic scale and compute the hybridization between the vacancy and the twisted pair of $\pi$-bands, showing a clear dichotomy between AA/BB and AB/BA regions. This is in stark contrast to the recent description of TBG as a topological Kondo lattice problem with suitably defined ``impurity'' limits (i.e., realizing local moments on the moir\'e scale)~\cite{song_magic-angle_2022, hu_kondo_2023, chou_kondo_2023, calugaru_tbg_2023, lau_topological_2023, zhou_kondo_2023, chou_scaling_2023, hu_symmetric_2023, shankar_kondo_2023}. Here, we focus on atomic-scale vacancies that arise in realistic experimental settings across a wide array of moir\'e materials (e.g., graphene~\cite{dietrich_vacancies_2023}, TMD~\cite{wang_atomic_2018, guo_moire_2021}, cuprate superconductors~\cite{pan_imaging_2000, alloul_defects_2009}).  
We use ab-initio derived vacancy states to construct an effective quantum impurity model for a realistic vacancy,
which is solved by combining the kernel polynomial method \cite{Weisse2006} and the numerical renormalization group \cite{Bulla2008} (KPM+NRG \cite{wu_aubry_2022}). 
Away from the magic angle, this realizes a tunable, pseudogap AIM where twisting the bilayers tunes
the vacancy through its quantum phase transition. 
At the magic angle, the impurity is always Kondo-screened at low temperatures. 
We study the distribution of Kondo temperatures $T_{\mathrm{K}}$ across the sample to show how $T_{\mathrm{K}}$ in the AB region is strongly suppressed relative to the AA region.

\begin{figure*}
    \centering
    \includegraphics[width = 0.99\textwidth]{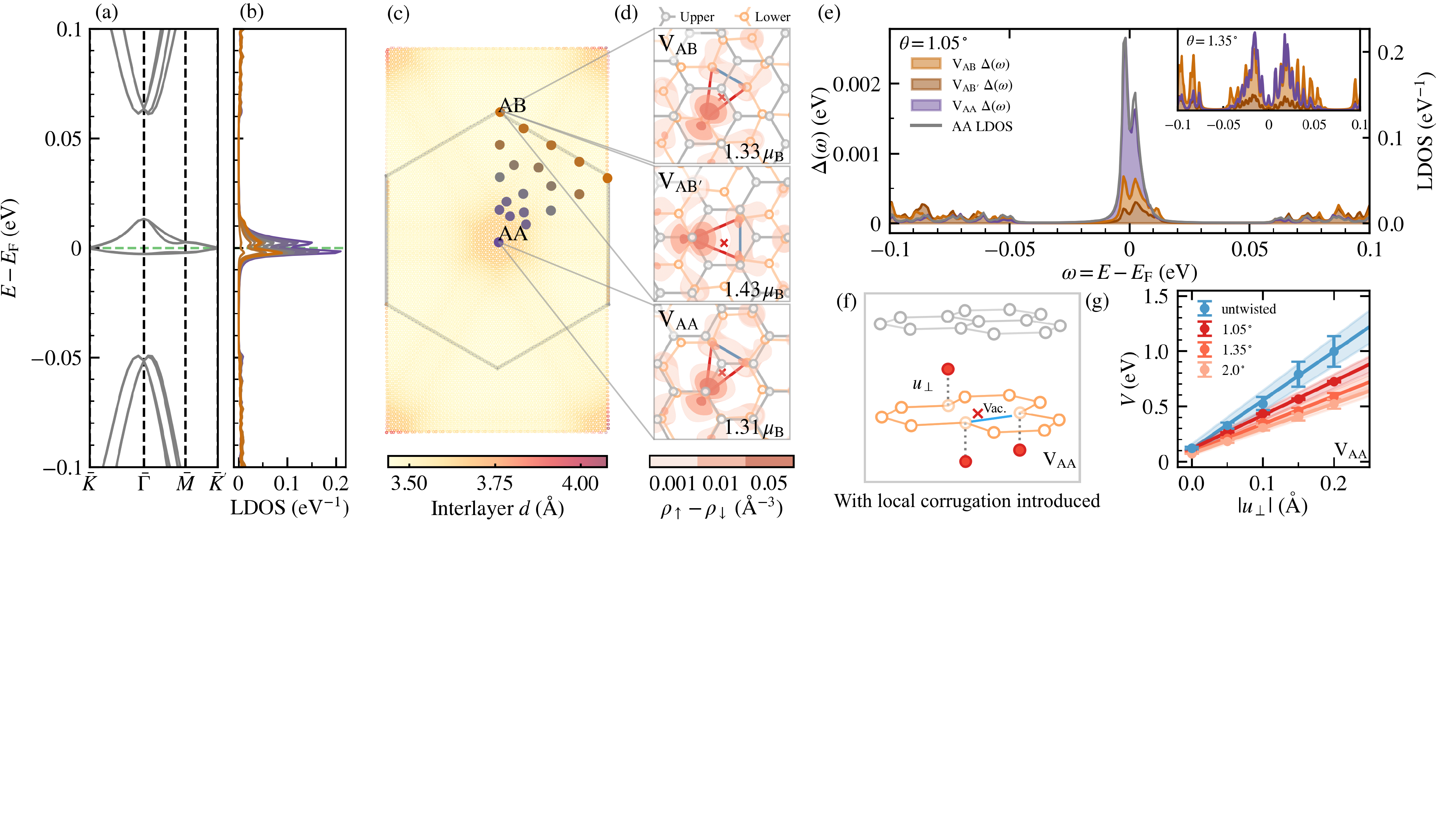}
    \caption{{\bf Atomic-scale modeling of TBG with a vacancy}.
    (a) Band structure of pristine TBG at $\theta_{\text{M}}$ without vacancy. $\bar{K}$ and $\bar{K}'$ indicate the two valleys of the moir\'e Brillouin zone. 
    (b) Projected local density of states (LDOS) at the potential vacancy sites denoted in (c), color-coded. 
    (c) The C atoms in the relaxed pristine TBG at $\theta_{\text{M}}$, with the center of AA and AB regions annotated. The atoms are colored-coded according to the local interlayer spacing. The hexagon represents the moir\'e unit cell.
    The 19 potential vacancy sites chosen for computing the LDOS in (b) are indicated by thicker dots. 
    (d) $\rho_{\uparrow}$\,$-$\,$\rho_{\downarrow}$ for three typical vacancy configurations, computed in untwisted bilayer graphene using DFT. $\mathrm{V_{AB}}$ and $\mathrm{V_{AB^\prime}}$ vacancies could be present in TBG AB regions, while $\mathrm{V_{AA}}$ vacancies are typical in AA regions.
    The grey (orange) circles denote the C atoms in the upper (lower) layers, and the red cross denotes the vacancy.
    The blue (red) lines highlight the shorter (longer) distances between three adjacent sites.
    The clouds of $\rho_{\uparrow}$\,$-$\,$\rho_{\downarrow}$ are plotted in red, showing that most of the excessive spin is contributed by the $\sigma$-lobe at the isolated adjacent site. 
    (e)  
    Hybridization function at $\theta_{\mathrm{M}}$ between the V$_\sigma$ state and the TBG bath for the three vacancy configurations.
    The LDOS for the AA site is also plotted in the grey curve for comparison. 
    The insets show the hybridization at $\theta \!=\! 1.35^\circ$, 
    which features a Dirac-cone-like low-energy dispersion with larger flat-band bandwidth compared to $1.05^\circ$.
    (f) Side view of AA-stacked bilayer graphene near the vacancy site, with manual vertical displacements of the three adjacent sites away from equilibrium, denoted by $u_{\perp}$.
    The bonded sites (connected by a blue line) are displaced downward by the same $u_{\perp}$.
    (g) 
    Hybridization
    strength $V$ between the V$_\sigma$ state and the bath versus $u_{\perp}$ for the untwisted V$_{\mathrm{AA}}$ vacancy (blue) and when coupled to the TBG bath (red).
    Lines are linear fits, 
    showing that $V$ is tunable by $u_{\perp}$ and $\theta$.
    }
    \label{fig:ldos}
\end{figure*}

{\it Microscopic picture.}---%
To set the stage, we investigate how the vacancy $\sigma$ (V$_\sigma$) state in TBG at the magic angle ($\theta_{\text{M}} \!=\! 1.05^\circ$~\cite{bistritzer2011moire}) hybridizes with the twisted pair of $\pi$-bands using an accurate machine-learned tight-binding model~\cite{pathak_accurate_2022}, combined with embedded V$_\sigma$ states from density functional theory (DFT) calculations~\cite{Supp}.
Figure~\ref{fig:ldos}(c) shows the structure of TBG at $\theta_{\text{M}}$ obtained by fully relaxing free-standing TBG using the interatomic potential model~\cite{brenner_second-generation_2002, kolmogorov_registry-dependent_2005, ouyang_nanoserpents_2018} with the method described in Ref.~\cite{rakib_corrugation-driven_2022, krongchon_registry-dependent_2023}, using the molecular dynamics (MD) simulation package LAMMPS~\cite{thompson_lammps_2022}. 
The in-plane and out-of-plane atomic relaxations manifest in enlarged AB/BA regions, consistent with previous work~\cite{dai_twisted_2016, zhang_structural_2018, lucignano_crucial_2019, yoo_atomic_2019, rakib_corrugation-driven_2022, krongchon_registry-dependent_2023}.
Figure~\ref{fig:ldos}(a) shows the band structure of fully relaxed pristine TBG in the atomic-scale tight-binding model, and Fig.~\ref{fig:ldos}(b) the local density of states (LDOS) projected onto the potential vacancy sites indicated in panel (c).
The flat-band states are mostly localized in the AA regions (except at $\Gamma$) and have decreasing projections onto sites further away from the AA center.
This indicates that an impurity in the AA (AB) regions hybridizes more (less) with the localized flat-band states.

To understand microscopically how vacancy states hybridize with the $\pi$-bands in TBG, we first consider a monovacancy in single-layer graphene.
Removing one atom leaves dangling vacancy V$_\pi$ and V$_\sigma$ orbitals at the three adjacent atoms, which undergo a Jahn--Teller distortion so that one isolated atom moves further away from the vacancy, and the other two atoms move closer to re-bond.
This leads to one V$_\sigma$ state localized at the isolated site near the Fermi level and a V$_\pi$ quasi-localized zero mode~\cite{palacios_critical_2012, nanda_electronic_2012}. 
In the experimentally relevant regime, the effect of the V$_\pi$ zero mode can be qualitatively captured by a renormalization of the V$_\sigma$ Coulomb interaction~\cite{jiang_inducing_2018}; we thus focus on the V$_\sigma$ state in the following.

In single-layer graphene, the coupling between the V$_\sigma$ state and the $\pi$-bands requires finite local corrugation that breaks the mirror symmetry~\cite{jiang_inducing_2018,may_modeling_2018}.
This coupling arises naturally if a second, untwisted layer is stacked onto free-standing graphene, which breaks the mirror symmetry near the vacancy site in the lower layer.
To capture this, we performed DFT calculations using a $6\!\times\! 6$ supercell of free-standing untwisted bilayer graphene with one vacancy in the bottom layer in three typical local environments~\cite{Supp}, named after the registry of the two sheets of graphene. 
Note that, in AB-stacked graphene, there are two types of vacancies, $\mathrm{V_{AB}}$ and $\mathrm{V_{AB'}}$ (see Fig.~\ref{fig:ldos}(d)).
Similar to single-layer graphene, the three adjacent atoms near the vacancy site relax to a final equilibrium configuration with almost no corrugation near the vacancy.
Figure~\ref{fig:ldos}(d) shows the calculated excessive spin density, $\rho_{\uparrow} \!-\! \rho_{\downarrow}$~\footnote{The spin $\uparrow$ or $\downarrow$, given by a spin-polarized calculation, do not correspond to any real-space direction.}, which is centered at the isolated adjacent C atom, contributed mainly by the $\sigma$-lobe toward the vacancy.
The details of the calculations and the electronic structure of bilayer graphene with a vacancy are summarized in~\cite{Supp}.

In
the dilute limit, we expect the $6\!\times\! 6$ supercell simulation of the V$_\sigma$ state in untwisted bilayer graphene to mimic the actual vacancy in TBG near $\theta_{\text{M}}$.
Using our DFT results in untwisted bilayer graphene, we compute the hybridization function between the dangling V$_\sigma$ state and the ``bath'' (TBG \textit{without} the vacancy and its three adjacent sites) from the microscopic model, 
$     \Delta_{\text{micro}}(\omega) \!=\! \pi\sum_{n, \mathbf{k}}|V_{n\mathbf{k}}|^2 
     \delta ( \omega - \epsilon_{n\mathbf{k}} ) $.
Here, $V_{n\mathbf{k}} \!=\! \braket{\phi_{\text{V}_\sigma} | H_{\text{V$_\sigma$-bath}} \mathcal{P}_{\text{bath-TBG}} |\psi_{n\mathbf{k}}}$ represents the tunneling matrix element between the V$_\sigma$ state $\ket{\phi_{\text{V}_\sigma}}$ and the pristine TBG eigenstate $\ket{\psi_{n\mathbf{k}}}$ with eigenvalue $\epsilon_{n\mathbf{k}}$.
$H_{\text{V}_\sigma\text{-bath}}$ is the hopping between the V$_\sigma$ state and the C atoms in the bath; $\mathcal{P}_{\text{bath-TBG}}$ projects the TBG eigenstate from the Hilbert space of $N$-site TBG to that of the $(N \!-\! 1)$-orbital $\pi$ bath.
Figure~\ref{fig:ldos}(e) shows $\Delta_{\text{micro}}(\omega)$
for the three vacancy 
environments at $\theta_{\text{M}}$ 
and
the LDOS at the AA center, with insets showing the result at $1.35^{\circ}$. 
In comparison,
V$_{\text{AA}}$ hybridizes much stronger with the bath, especially with the flat-band states.

In experiments, the substrate almost always induces 
local corrugation in TBG, which is not captured in our MD simulations. 
Therefore, we manually introduced local corrugation near the vacancy to study how it changes the hybridization. 
Figure~\ref{fig:ldos}(f) shows a schematic side view of how the three vacancy-adjacent atoms are displaced by $u_{\perp}$ either upward or downward~\cite{Supp} for a V$_{\mathrm{AA}}$ vacancy.
Figure~\ref{fig:ldos}(g) shows that the hybridization strength $V$ increases with $u_{\perp}$ and decreases with 
$\theta$.
For $u_{\perp} \!>\! 0.2$\,\AA,
a localized V$_{\sigma}$ state can no longer be identified.
We find that V$_{\text{AA}}$ vacancies are more sensitive to twisting~\cite{Supp}.

In summary, a vacancy in bilayer graphene induces a localized spin density, 
which hybridizes with the $\pi$-bath with a strength tunable via the local environment, atomic corrugation, and the twist angle, suggesting that TBG with a monovacancy realizes a tunable, pseudogap AIM. 
To make the AIM tractable, we 
construct
an effective 
model with the 
impurity parameters
derived from the microscopic model and the hybridization function 
from
a simpler TBG bath. The latter has all of the salient features we have just found and allows us to describe the spectral properties of TBG down to sufficiently low energy scales to treat the Kondo effect accurately.

\begin{figure}
\centering
     \includegraphics[width=0.48\textwidth]{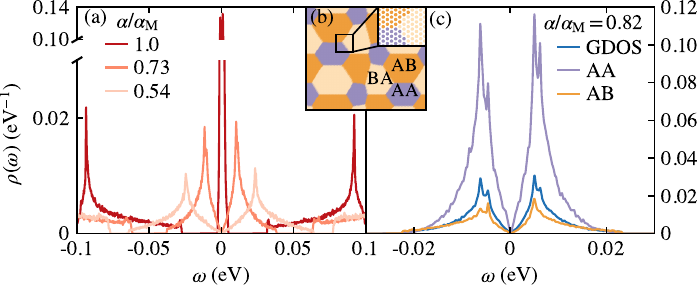}
    \caption{{\bf Hybridization functions for the effective lattice model of TBG} in Eq.~\eqref{eqn:model-hyb}. (a) The model GDOS at different twist parameters, where $\alpha \!=\! 0.081$ ($w \!=\! 0.11 \,\mathrm{eV}$) corresponds to the magic angle. 
    (b) Tunneling map, where different colors indicate the locally dominating tunneling according to the effective lattice model \cite{Supp}. 
    The inset shows the microscopic lattice spacing across a patch. (c) Comparison of GDOS and LDOS in the AA and AB regions at $\alpha \!=\! 0.067$ ($w \!=\! 0.09 \,\mathrm{eV}$). The LDOS is averaged over 200 samples of random twisted boundary conditions; the GDOS is additionally averaged over the origin of rotation in TBG across 400 samples. 
}
\label{fig:dos}
\end{figure}

{\it Quantum impurity model.}---%
We use the AIM, with the Hamiltonian 
$H \!=\! H_\text{host} \!+\! H_\text{hyb} \!+\! H_\text{imp}$,
to emulate TBG with a vacancy. 
The host contributions are typically written in the single-particle eigenbasis, where $\epsilon_k$ is the eigenenergy of a state created by $c^\dag_{k\sigma}$ with a wavefunction $\phi_{k\sigma}(j) = \langle j,\sigma | \epsilon_k\rangle$ at lattice site $j$~\footnote{Here, $k$ labels the eigenstate, but does not represent the momentum. The lattice model is defined on an incommensurate lattice. Therefore, momentum is not a well-defined quantum number.}. Then,
\begin{equation}
H_\text{host} \!=\! \sum_{k,\sigma} \epsilon_k c_{k\sigma}^\dag c_{k\sigma} 
, \quad
H_\text{hyb} \!=\! V \sum_{\sigma} \bigl( d_{\sigma}^\dag c_{R\sigma} \!+\! \mathrm{H.c.} \bigr),
\label{eqn:model-hyb}
\end{equation}
where $R$ labels the impurity site, $c_{R\sigma} \!=\! \sum_k \phi_k(R) c_{k\sigma}$, and $V \!>\! 0$ is the hybridization strength.
The effect of the host on the impurity is described by the hybridization function
\begin{equation}
\label{eq:hyb}
    \Delta_R(\omega)
    = \pi V^2 \sum_k |\phi_k(R)|^2 \delta(\omega-\epsilon_k)
    \equiv \pi V^2 \rho_R(\omega),
\end{equation}
with $\rho_R$ the host LDOS per spin orientation ($\phi_k\equiv \phi_{k\sigma}$). 
One approximation we consider to help gain physical insight into the problem ignores the spatial contribution of the wavefunction to the LDOS, so that the LDOS in the hybridization function is replaced by the global DOS (GDOS, per spin orientation, per lattice site) $\rho(\omega)=N^{-1}\sum_k \delta(\omega - \epsilon_k)$. 

To describe the $\omega$-dependence of $\Delta_{R}(\omega)$, we use a microscopic lattice model of TBG \cite{PhysRevResearch.2.023325}
derived from the Bistritzer--MacDonald (BM) continuum model \cite{bistritzer2011moire,Supp}, which can be scaled up in system size, 
provides
higher resolution for the ultra-low-energy features in the LDOS and captures the emergent multifractality in TBG's wavefunctions at $\theta_{\text{M}}$ \cite{Supp}.
We can modify 
$\Delta_{R}(\omega)$ by varying either
the twist angle $\theta$ or the interlayer tunneling $w$,
since only their ratio matters at small twist angles in the form $\alpha \!\equiv \! w/[2 v_{\mathrm{F}} k_{\mathrm{D}} \sin(\theta/2)]$. Here, the Fermi velocity is $v_{\mathrm{F}} \!=\! 3 ta_0/(2\hbar)$ with $t \!=\! 2.8 \, \mathrm{eV}$,
and the distance from the $\Gamma$ to the Dirac point is $k_{\mathrm{D}} \!=\! 4\pi/(3a_0)$ with $a_0 \!\approx\! 2.46\,\mathrm{\r{A}}$ the graphene lattice constant.
It is more convenient for us to vary $w$ at fixed $\theta \!=\! 1.05^{\circ}$ as we treat the incommensurate twist via a rational approximation. The magic angle $\alpha_{\mathrm{M}}$ then occurs at $w \!=\! 0.11 \,\mathrm{eV}$.

We focus on the charge neutrality point and show
the DOS of the TBG lattice model in Fig.~\ref{fig:dos}(a).
Relaxation in the lattice model is accounted for by breaking the symmetry in the tunneling between the 
AA and AB regions,
with $w_{\text{AA}}/w_{\text{AB}} \!=\! 0.75$~\cite{PhysRevResearch.1.013001}.
In the BM model, the GDOS can have a charge neutrality van Hove singularity at the magic angle, which  
(ignoring how the impurity is embedded in the host) leads to Kondo screening of the impurity~\cite{shankar_kondo_2023}.
Incorporating the impurity location,
the tunneling strengths in the lattice model mark different sublattice tunneling 
geometries
dominated by AA/BB or AB/BA 
regions (Fig.~\ref{fig:dos}(b)). The LDOS in each representative region, depicted in Fig.~\ref{fig:dos}(c) away from 
$\alpha_{\mathrm{M}}$, 
reflects how the probability density of wavefunctions in the miniband are concentrated near the AA sites, consistent with our atomic-scale model results~\cite{Supp} and expectations from previous TBG Wannier-function calculations~\cite{koshino_maximally_2018, kang_symmetry_2018, po_faithful_2019, carr_derivation_2019, carr_electronic-structure_2020}.

Finally, the impurity part in the Hamiltonian reads
\begin{equation}
\label{eq:H_imp}
    H_\text{imp} = \epsilon_d(\hat{n}_{d\uparrow} + \hat{n}_{d\downarrow}) + U \hat{n}_{d\uparrow} \hat{n}_{d\downarrow}.
\end{equation}
An impurity state with spin $\sigma$ and on-site repulsion $U$, localized at the vacancy site $R$, is created by $d^\dag_\sigma$, has a number operator $\hat{n}_{d\sigma} \!=\! d_\sigma^\dagger d_\sigma$, and an energy $\epsilon_d$ measured from the host Fermi energy $E_{\mathrm{F}} \!=\! 0$.
We choose $U \!=\! 2.2\,\mathrm{eV}$ and $\epsilon_d \!=\! -0.5\,\mathrm{eV}$ as motivated by our microscopic analysis~\cite{Supp}.
Note that the hybridization functions already break particle-hole symmetry and that the TBG half bandwidth (of the full spectrum, not only the miniband) $D$ depends weakly on $\alpha$, $D(\alpha) \!\approx\! 8 \, \mathrm{eV}$.
Below, the hybridization strength is either varied or chosen as $V\!=\! 1\,\mathrm{eV}$ to estimate the typical $T_{\mathrm{K}}$.
The effective Kondo coupling is $J \!\sim\! V^2/U$.

\begin{figure}[t!]
    \centering
    \includegraphics[width = 0.49\textwidth]{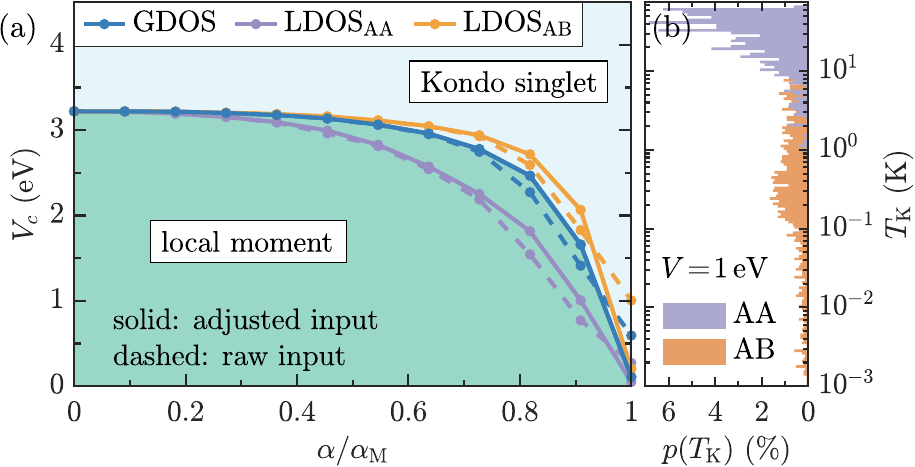}
	\caption{{\bf Solution of the quantum impurity model.} (a) Phase diagram at $T \!=\! 0$. The critical hybridization strength $V_c$ that separates the local-moment ($\mu_{\mathrm{eff}} \!=\! 1/4$) and Kondo-singlet phase ($\mu_{\mathrm{eff}} \!=\! 0$) as a function of $\alpha = \! w/[2 v_{\mathrm{F}} k_{\mathrm{D}} \sin(\theta/2)]$ vanishes linearly at the magic angle $\alpha_{\mathrm{M}} \!\approx\! 1/\sqrt{3}$ (cf.\ Eq.~\eqref{eq:critical_coupling}). Dashed lines show results for $\Delta \!\propto\! \rho$ with the GDOS $\rho$ or with an LDOS $\rho_R$ ($R$ in the AA or AB region), see Eq.~\eqref{eq:hyb}. For the solid lines, the input is adjusted to match the analytically known asymptotic low-energy behavior \cite{Supp}.
    (b) Distribution of $T_{\mathrm{K}}$ for different impurity locations across magic-angle TBG. We consider $V \!=\! 1\,\mathrm{eV}$ and roughly 500 (1500) sites for the AA (AB) region. The AB, compared to the AA distribution, is broader, centered at a smaller value, and has a tail reaching down to very low $T_{\mathrm{K}}$.
    }
    \label{fig:phase_diagram_gdos}
\end{figure}

{\it Many-body solution.}---%
We first 
consider
the $T\!=\! 0$ phase diagram for a vacancy in TBG with variable hybridization strength $V$ and twist parameter $\alpha$.
A single sheet of graphene realizes a clear-cut pseudogap AIM
with a local-moment and Kondo-singlet phase.
Away from  
$\alpha_{\mathrm{M}}$,
the GDOS of TBG still follows the pseudogap behavior: at low energies,
$\rho(\omega) \!\sim\! |\omega|/v^2$, where $v \!=\! v(\alpha)$ is the renormalized Dirac velocity. 
As $v(\alpha)$ vanishes at 
$\alpha_{\mathrm{M}}$
($\alpha_{\mathrm{M}} \!\approx\! 1/\sqrt{3}$ for the first magic angle in \cite{bistritzer2011moire}), we may expand it as $v(\alpha) \!\sim\! |\alpha-\alpha_{\mathrm{M}}|$ close to $\alpha_{\mathrm{M}}$. From previous NRG studies, we know that such a particle-hole asymmetric pseudogap AIM has a critical Kondo coupling $J_c$ with $\rho_0 J_c \!\sim\! \mathcal{O}(1)$~\cite{chen1995kondo, ingersent_critical_2002, PhysRevB.70.214427, PhysRevB.57.14254, fritz2013physics}.
Hence, from $\rho_0\sim 1/v^2(\alpha)$, we expect
\begin{equation}
J_c \sim |\alpha-\alpha_{\mathrm{M}}|^2
\quad \Leftrightarrow \quad 
V_c \sim |\alpha-\alpha_{\mathrm{M}}|.
\label{eq:critical_coupling}
\end{equation}
At 
$\alpha_{\mathrm{M}}$,
$\rho(\omega)$ is smooth at low $\omega$ with finite $\rho(0)$. Hence, at $T \!=\! 0$, a Kondo-singlet phase is found for arbitrarily small $V$.

We use KPM with a linear lattice size $L \!=\! 569 a_0$ \cite{PhysRevResearch.2.023325} and an expansion order of $N_C \!=\! 2^{18}$ to calculate the Wilson-chain coefficients. 
Then, from the NRG impurity contribution to the spin susceptibility $\chi_{\mathrm{imp}}$, we extract the effective magnetic moment 
$\mu_{\mathrm{eff}} \!=\! \lim_{T\to 0} T \chi_{\mathrm{imp}}(T)$.
This yields a phase diagram, Fig.~\ref{fig:phase_diagram_gdos}(a), where
the local-moment and Kondo-singlet phases are characterized by $\mu_{\mathrm{eff}} \!=\! 1/4$
and $\mu_{\mathrm{eff}} \!=\! 0$, respectively.
$V_c(\alpha)$ for hybridization functions proportional to an LDOS 
(with $R$ 
in either the AA or AB region) and to the GDOS (equivalent to the 
$R$-averaged LDOS)
behave qualitatively similarly. As expected, in the AA region, the enhanced LDOS leads to a smaller $V_c$ relative to the AB/BA regions.
For the GDOS, we  
we can use
the asymptotic low-energy behavior mentioned before ($\rho(\omega)|_{\alpha\neq\alpha_{\mathrm{M}}} \!\sim\! |\omega|$, $\rho(\omega)|_{\alpha_{\mathrm{M}}} \!=\! \mathrm{const}$)
to adjust and extend the Wilson-chain input at energy scales below the KPM resolution 
\cite{Supp}.
The data with adjusted input nicely reproduces Eq.~\eqref{eq:critical_coupling} and confirms our expectations: There is a finite $V_c$ for all $\alpha \!\neq\! \alpha_{\mathrm{M}}$, decreasing linearly with $\alpha$ close to $\alpha_{\mathrm{M}}$.
At $\alpha \!=\! \alpha_{\mathrm{M}}$, any $V \!>\! 0$ leads to Kondo screening, 
consistent with 
a recent study using 
the GDOS in TBG~\cite{shankar_kondo_2023}.

Focusing on $\alpha_{\mathrm{M}}$, where the ground state is always Kondo-screened, we may ask below which $T$ the Kondo-singlet phase occurs across the 
sample, i.e., 
how $T_{\mathrm{K}}$ changes with the location of the impurity at fixed 
$V$ (say, $1\,\mathrm{eV}$).
We use $T_{\mathrm{K}} \!\simeq\! 1/|4\chi_{\mathrm{loc}}|$ \cite{Filippone2018}
as a robust and efficient estimate, where 
$\chi_{\mathrm{loc}} \!=\! \partial_h \langle S^z \rangle |_{h=0}$ (with $\langle S^z \rangle$ the local magnetization) is the local susceptibility computed at $T\!=\! 0$. 
In Fig.~\ref{fig:phase_diagram_gdos}(b), we plot the distribution of Kondo scales found for a large number of different sites.
There is a strict dichotomy between vacancies in the AA versus AB regions. 
The LDOS throughout the AA region is rather similar, leading to a narrow distribution of Kondo scales.
By contrast, the LDOS in the AB region varies widely at low energies (as it includes $\text{V}_{\text{AB}}$ and $\text{V}_{\text{AB}^\prime}$ contributions) and is generally smaller than in the AA region. This leads to a broad distribution of Kondo scales, centered at a smaller value than in the AA region, and with a tail to vanishing $T_{\mathrm{K}}$. 
This tail is a signature of a broad distribution we expect~\cite{kettemann_kondo_2012, gammag_distribution_2016, slevin_multifractality_2019, wu_aubry_2022} to result from the multifractal wavefunctions at the magic angle that arise in several incommensurate models of TBG~\cite{fu2020magic,gonccalves2021incommensurability} including $H_{\text{host}}$ in Eq.~\eqref{eqn:model-hyb} (see \cite{Supp}).

In reality, the TBG bath 
becomes correlated
very close
to 
$\theta_{\mathrm{M}}$,
and the non-interacting bath description breaks down. 
Therefore, at $T\!=\! 0$, our results are directly applicable across the majority of the phase diagram in Fig.~\ref{fig:phase_diagram_gdos}(a), 
while
future work is needed to incorporate the strongly correlated bath that can induce a gap at charge neutrality~\cite{lu_superconductors_2019}.
At finite temperatures above the correlated gap, our results serve as a description of the normal state of the defect-induced Kondo effect in TBG.

{\it Conclusion.}---%
Using ab-initio calculations, we embedded a vacancy into pristine TBG and demonstrated how it hybridizes with the low-energy miniband in the vicinity of the magic angle. From this insight, we built an effective AIM that we solved with KPM+NRG~\cite{wu_aubry_2022}. We found a variety of many-body ground states and a pseudogap quantum critical point tunable by the twist angle. At the magic angle, the vacancy is always Kondo-screened, leading to a 
distribution of Kondo temperatures
that is broad in the AB region due to the underlying multifractal single-particle eigenstates \cite{fu2020magic,gonccalves2021incommensurability}.
We propose the STM response of such Kondo-induced vacancy states as a probe of the underlying many-body ground states in TBG and 
moir\'e materials more broadly.

{\it Acknowledgments.}---%
We thank Kevin Ingersent and Justin Wilson for insightful discussions and collaborations on related work.
Y.C.\ thanks Tawfiqur Rakib for help in setting up MD simulations for TBG, and Shang Ren and Michele Pizzochero for valuable discussions.
This work has been supported in part by the NSF CAREER Grant No.~DMR 1941569 (A.K.W., J.H.P.), the BSF Grant No.~2020264 (J.Y., J.H.P.), the Alfred P.~Sloan Foundation 
through a Sloan Research Fellowship (J.H.P.), 
Department of Energy DOE-FG02-99ER45742 (E.Y.A.),
the Gordon and Betty Moore Foundation EPiQS initiative GBMF9453 (E.Y.A.), the NSF Grant No.~DMR 1954856 (D.V.), and the U.S. Department of Energy, Office of Science, Office of Advanced Scientific Computing Research, and Office of Basic Energy Sciences, Scientific Discovery through Advanced Computing (SciDAC) program (G.K.).
F.B.K.\ acknowledges support by the Alexander von Humboldt Foundation through the Feodor Lynen Fellowship.
Y.C.\ acknowledges support from the Abrahams Postdoctoral Fellowship of the Center for Materials Theory at Rutgers University.
The KPM computations were performed using the Beowulf cluster at the Department of Physics and Astronomy of Rutgers University and the Amarel cluster provided by the Office of Advanced Research Computing (OARC) \footnote{http://oarc.rutgers.edu} at Rutgers, The State University of New Jersey.
The NRG results were obtained using the QSpace tensor library developed by A.~Weichselbaum 
\cite{Weichselbaum2012a,*Weichselbaum2012b,*Weichselbaum2020}
and the NRG toolbox by Seung-Sup B.~Lee \cite{Lee2016,Lee2017,Lee2021}.
The Flatiron Institute is a division of the Simons Foundation.

% Combined references
 
\end{document}

\title{Supplementary information of ``Vacancy induced tunable Kondo effect in twisted bilayer graphene''}
\author{Yueqing Chang}
 \affiliation{Department of Physics and Astronomy,   
Rutgers University, Piscataway, NJ 08854, USA}
  \affiliation{Center for Materials Theory,  
Rutgers University, Piscataway, NJ 08854, USA}
\author{Jinjing Yi}
 \affiliation{Department of Physics and Astronomy,   
Rutgers University, Piscataway, NJ 08854, USA}
  \affiliation{Center for Materials Theory,  
Rutgers University, Piscataway, NJ 08854, USA}
\author{Ang-Kun Wu}
  \affiliation{Department of Physics and Astronomy,   
Rutgers University, Piscataway, NJ 08854, USA}
  \affiliation{Center for Materials Theory,  
Rutgers University, Piscataway, NJ 08854, USA}
\author{Fabian B.~Kugler}
\affiliation{Center for Computational Quantum Physics, Flatiron Institute, 162 5th Avenue, New York, NY 10010, USA}
\affiliation{Department of Physics and Astronomy,   
Rutgers University, Piscataway, NJ 08854, USA}
  \affiliation{Center for Materials Theory,  
Rutgers University, Piscataway, NJ 08854, USA}
\author{Eva Andrei}
 \affiliation{Department of Physics and Astronomy, 
Rutgers University, Piscataway, NJ 08854, USA}
\author{David Vanderbilt}
\affiliation{Department of Physics and Astronomy,   
Rutgers University, Piscataway, NJ 08854, USA}
  \affiliation{Center for Materials Theory,  
Rutgers University, Piscataway, NJ 08854, USA}
\author{Gabriel Kotliar}
\affiliation{Department of Physics and Astronomy,   
Rutgers University, Piscataway, NJ 08854, USA}
  \affiliation{Center for Materials Theory,  
Rutgers University, Piscataway, NJ 08854, USA}
\affiliation{Condensed Matter Physics and Materials Science Department,\looseness=-1\,  
Brookhaven National Laboratory, Upton, NY 11973, USA}
\author{J.~H.~Pixley}
\affiliation{Department of Physics and Astronomy,   
Rutgers University, Piscataway, NJ 08854, USA}
  \affiliation{Center for Materials Theory,  
Rutgers University, Piscataway, NJ 08854, USA}
\affiliation{Center for Computational Quantum Physics, Flatiron Institute, 162 5th Avenue, New York, NY 10010, USA}

\maketitle

\tableofcontents

\section{Ab-initio study of monovacancy in twisted bilayer graphene}

\subsection{DFT simulation of the monovacancy in untwisted bilayer graphene}
\label{suppsec:DFT_monovacancy_in_bl}

\begin{figure}
    \centering
    \includegraphics[width=0.95\textwidth]{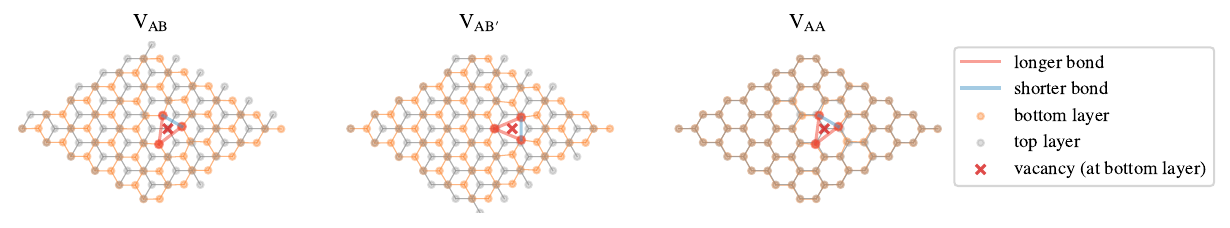}
    \caption{Relaxed lattice structure of the $6\times 6$ supercell of the bilayer graphene with vacancy systems in three different local environments, named after stacking registry.
    The vacancy is placed on the bottom layer, denoted by the red cross. 
    The distorted isosceles triangle formed by the three nearest neighbors of the vacancy is denoted by red dots, with the red and blue lines annotating the longer and shorter bonds among them.}
    \label{fig:bl_structure}
\end{figure}

To study how local atomic environments affect the vacancy resonant states and take into account the V$_\sigma$ states localized at the nearest neighbor sites, which occur near the Fermi level, we performed density functional theory (DFT) calculation of bilayer graphene with two different registries and three different vacancy configurations: V$_{\text{AA}}$, V$_{\text{AB}}$, V$_{\text{AB}^\prime}$.
For each bilayer structure, we constructed a 143-atom 6$\,\times\,$6 supercell with a single vacancy at the bottom layer. 
The lattice relaxation and self-consistent-field calculations were performed using the Vienna Ab initio Simulation Package~\cite{kresse_ab_1993, kresse_efficiency_1996, kresse_efficient_1996}, with a fully spin-polarized Perdew-Burke-Ernzerhof exchange-correlation functional~\cite{perdew_generalized_1996} with a 6\,$\times$\,6\,$\times$\,1 $k$-grid, with a force convergence tolerance 0.05\,eV/\AA. 
The fully relaxed structures of the three different vacancy configurations exhibit a similar Jahn-Teller distortion near the vacancy, as shown in Fig.~\ref{fig:bl_structure}.
The distortions show that one nearest-neighbor carbon atom moves away from the vacancy site, while the other two atoms move closer toward each other to re-bond.
The equilibrium bond lengths $l_1$, $l_1$ and $l_2$ satisfy $l_1>l_0$ and $l_2<l_0$, where $l_0$ is the equilibrium next-nearest-neighbor distance in the pristine single layer graphene. 

\begin{figure}
    \centering
    \includegraphics[width=0.4\textwidth]{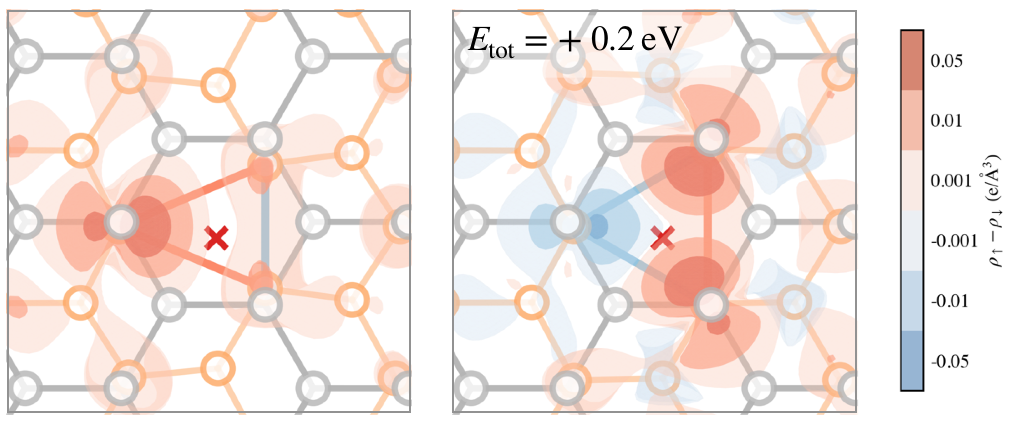}
    \caption{The two V$_{\text{AB}'}$ distortions that are close in energies. 
    The spin density $\rho_{\uparrow} - \rho_{\downarrow}$ isosurfaces are plotted for each configuration.
    We found that one can converge to the correct equilibrium configuration only with a finer $k$-grid (such as $6$\,$\times$\,$6$\,$\times$\,$1$) and a tighter force convergence tolerance.}
    \label{fig:v_ab2_configs}
\end{figure}

We also noticed that the relaxation of V$_{\text{AB}^\prime}$ takes a particularly long time since the potential energy surface of this type of vacancy is rougher and requires more time to resolve the lowest energy configuration.
We plotted the two configurations that are very close in energy for V$_{\text{AB}^\prime}$ in Fig.~\ref{fig:v_ab2_configs}.
We noticed that one can converge to the correct equilibrium configuration only with a finer $k$-grid (such as 6\,$\times$\,6\,$\times$\,1) and a tighter force convergence tolerance.
For the triangle formed by the three nearest-neighbor sites in V$_{\text{AB}^\prime}$, there is no carbon site at the top, while for the V$_{\text{AA}}$ or V$_{\text{AB}}$, the carbon site at the top of the triangle repels the lone $\sigma$ and $\pi$ electrons.
Therefore, for V$_{\text{AB}^\prime}$, another possible low-energy configuration (Fig.~\ref{fig:v_ab2_configs} right panel) is when two atoms come further, and one atom moves closer to the vacancy, leaving the total spin moment to be localized around two adjacent sites.
While for V$_{\text{AA}}$ and V$_{\text{AB}}$, repelled by the center-top carbon, two nearest-neighbor carbon sites always move closer to reform a 2-carbon bound state, significantly lowers the total energy and leads to faster convergence.

We then performed fully spin-polarized self-consistent-field (SCF) calculations on these four structures using a uniform 3\,$\times$\,3\,$\times$\,1  Monkhorst-Pack $k$-grid.
The formation energies of these three types of vacancies, computed as $E(N-1)-\frac{N-1}{N}E(N)$ are listed in Table~\ref{table:formation_energy_spin_moment}, comparable with the $7.0\pm 0.5\,$eV formation energy of a monovacancy in single-lager graphene as experimentally measured~\cite{thrower_point_1978}. 
For all three types of vacancies, the spin density $\rho_{\uparrow}-\rho_{\downarrow}$ localizes mainly on the further-away carbon, contributed mostly by the lone $\sigma$-electron, and partially by the lone $\pi$-electrons. 
The V$_\sigma$ state and the zero-mode V$_\pi$ state are occupied in the spin-up channel and empty in the spin-down channel, both contributing to the $S=1$ ground state spin density localized around the three nearest-neighbor sites.
Because of screening between the V$_\pi$ state and the $\pi$ bath, the V$_\pi$ state's contribution to the final spin magnetic moment is about $0.4\,\mu_{\mathrm{B}}$, resulting in a total spin magnetic moment $\sim 1.4\,\mu_{\mathrm{B}}$ (Table~\ref{table:formation_energy_spin_moment}).
In comparison, the spin magnetic moment of the monovacancy in single-layer graphene ranges from $1.04$ to $1.84\, \mu_{\mathrm{B}}$~\cite{nanda_electronic_2012}.
However, we point out that this discrepancy in $\mu_{\text{tot}}$ is likely due to a non-converged vertical displacement, sparse $k$-grid, and small supercell (high defect concentration)~\cite{palacios_critical_2012}.
Our calculation using a 12\,$\times$\,12\,$\times$\,1 $k$-grid yielded 1.23, 1.36, and 1.21\,$\mu_{\mathrm{B}}$ for the three vacancy types listed, slightly smaller than the values given by 6$\,\times\,$6$\,\times\,$1 grid listed in Table~\ref{table:formation_energy_spin_moment}.
Also, note that the dispersion in the $\text{V}_{\sigma}$ state might be greatly overestimated due to the vacancy-vacancy interaction introduced by the periodic boundary condition we used in DFT calculations.
This might lead to an underestimated vacancy magnetic moment~\cite{leccese_anomalous_2023}, which should have a value that is closer to 2\,$\mu_{\text{B}}$ than $\sim\,1.4\,\mu_{\text{B}}$.

In addition, one can estimate the intra-band interaction $U$ for the V$_\sigma$ state. 
Assuming that the Hubbard model of an isolated impurity applies, the single-particle energies, $\sim$\,$-0.5$ and $1.8$\,eV, of the V$_\sigma$ (occupied, spin up) and lowest unoccupied spin down V$_\sigma$ states would be $\epsilon$ and $\epsilon+U$, i.e., the energies needed for the removing one electron, $\ket{\sigma}\rightarrow \ket{0}$ ($\sigma=\uparrow, \downarrow$) and adding one electron, $\ket{\sigma} \rightarrow \ket{\uparrow\downarrow}$.
Here, $\ket{0}$, $\ket{\sigma}$, $\ket{\uparrow\downarrow}$ are the eigenstates of the isolated one-band impurity model, with on-site energy $\epsilon$ and Hubbard repulsion $U$.
This allows us to estimate the value of intra-orbital Hubbard interaction $U$, which is listed in Table~\ref{table:formation_energy_spin_moment}, in rough agreement with the value of 2.0\,eV given by Ref.~\cite{miranda_coulomb_2016}.
The value is smaller than the intra-orbital onsite Hubbard repulsion computed to be 4.4\,eV using the state-of-the-art constrained random phase approximations based on DFT results~\cite{schuler_optimal_2013}, and 3.6\,eV given by fixed-node diffusion Monte Carlo~\cite{changlani_density-matrix_2015} for carbon's $p_z$ orbitals.

\begin{center}
\begin{table}
\renewcommand{\arraystretch}{1.4}
\begin{tabular}{ m{7cm} m{3cm} m{3cm} m{3cm}}
  &  V$_{\text{AB}}$ & V$_{\mathrm{AB^\prime}}$ & V$_{\text{AA}}$\\[0.1cm]
 \hline
 Formation energy (eV)  & 5.66 & 5.64 & 5.48\\ 
 Total spin magnetic moment ($\mu_{\text{B}}$), $k$: 6$\times$6$\times$1 & 1.33 & 1.43 & 1.31 \\
 (contribution of V$_\sigma$ and V$_\pi$ ($\mu_{\text{B}}$)) & 1.0     ~~~~0.33 & 1.0   ~~~~ 0.43  & 1.0   ~~~~ 0.31 \\
 \hline
 V$_{\sigma}$ on-site energy $\epsilon$ (eV) & -0.45  & -0.48 & -0.51 \\ 
 V$_{\sigma}$ on-site Hubbard $U$ (eV) & 2.26  & 2.21 & 2.24 \\
V$_{\sigma}$-$\pi$ bath hybridization strength $V$ (eV) & 0.24(3) & 0.13(1) & 0.12(1) \\ % need to compute these values for untwisted bilayer
(contribution from top layer $V$ (eV) )& 0.12(1) & 0.09(1) & 0.07(1) \\ 
\end{tabular}
\caption{Computed properties of the three vacancy configurations in equilibirum, along with the estimated intra-orbital Hubbard $U$ and on-site energy $\epsilon$ for the V$_{\sigma}$ state and the coarse-grained hybridization $V$ between the V$_{\sigma}$ and the $\pi$ bath, in the untwisted bilayer graphene.}
\label{table:formation_energy_spin_moment}
\end{table}
\end{center}

\begin{figure}
    \centering
    \includegraphics[width=0.95\textwidth]{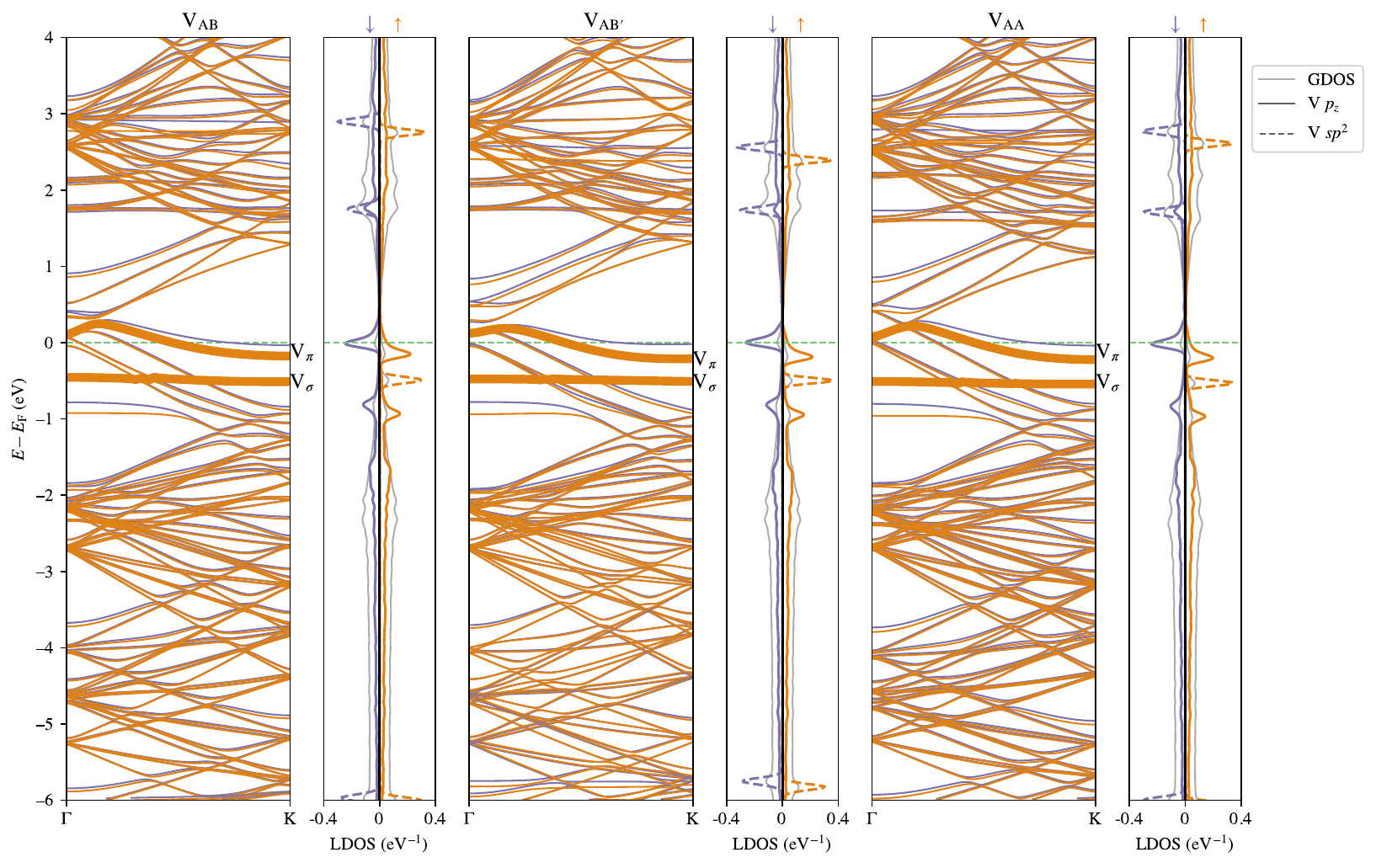}
    \caption{The bands and local density of states (LDOS) as computed by the tight-binding model derived from Wannierized DFT results for the three vacancy configurations in untwisted bilayer graphene.
    The two vacancy states of interest, V$_\pi$ and V$_\sigma$, are highlighted by the thick lines on the band structure plot.}
    \label{fig:bands}
\end{figure}

\begin{figure}
    \centering
    \includegraphics[width=0.45\textwidth]{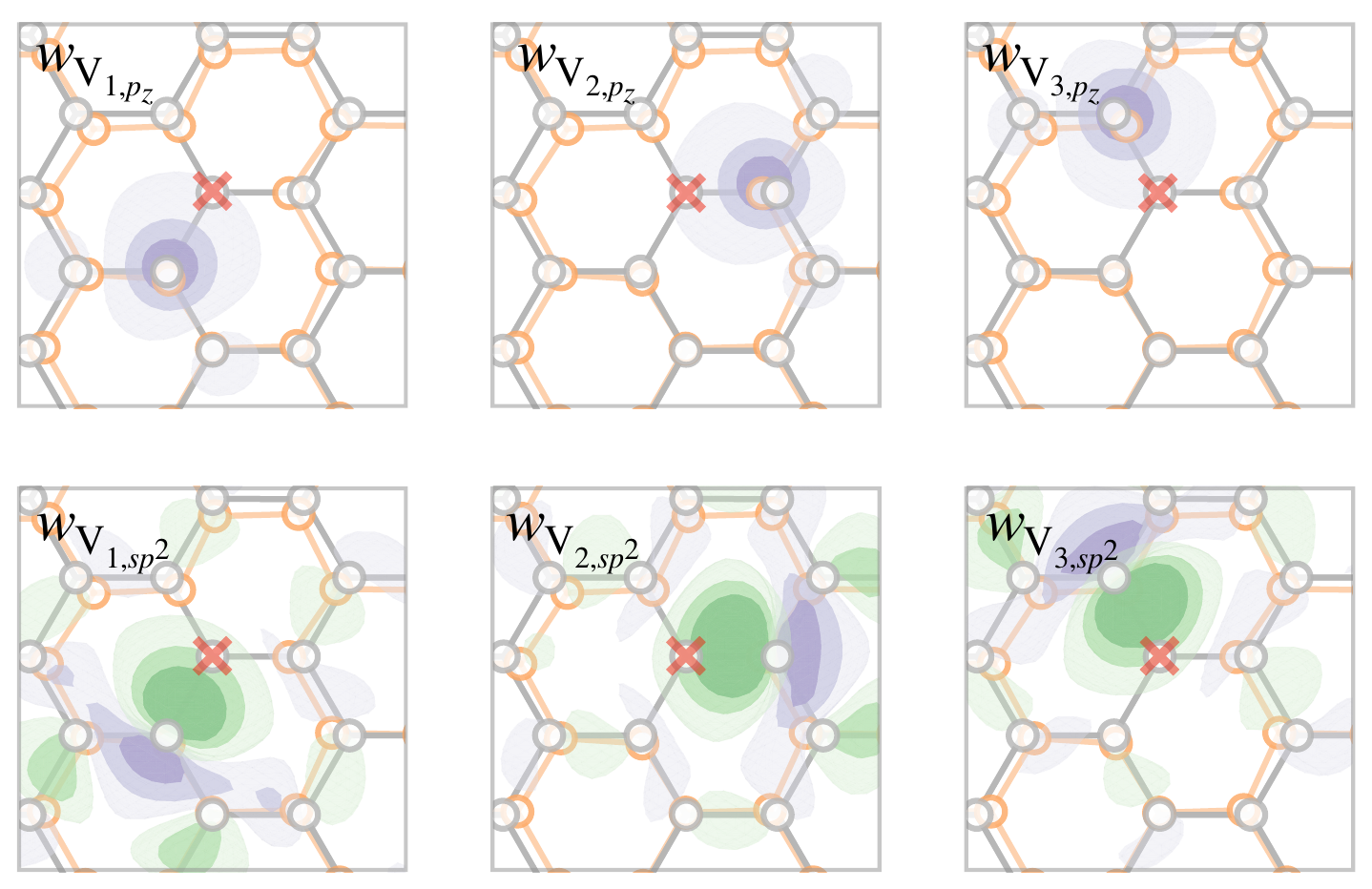}
    \caption{The rotated maximally localized Wannier functions (MLWFs) at the three adjacent sites of the vacancy in the V$_{\mathrm{AA}}$ configuration.
    The red cross denotes the location of the vacancy site at the bottom layer.}
    \label{fig:mlwfs}
\end{figure}

To study the hybridization of the vacancy states with the itinerant bands, we disentangled the Kohn-Sham wavefunctions near the Fermi level within a $\sim 20$\,eV window and downfolded to a set of maximally-localized Wannier functions (MLWFs)~\cite{marzari_maximally_1997} using the Wannier90 package~\cite{mostofi_wannier90_2008, mostofi_updated_2014, pizzi_wannier90_2020}.
The initial guess of the projections includes atomic $p_z$ orbitals on each carbon site and three $sp^2$ orbitals, with their lobes directed towards the vacancy site on the three vacancy nearest-neighbor sites.
The converged MLWFs at the three nearest neighbor sites are usually linear superpositions of each site's $p_z$ and $sp^2$ orbitals. 
Therefore, we performed a unitary rotation within each 2$\times$2 block such that the two onsite $sp^2$ and $p_z$ orbitals have zero hopping.
Then, we defined the orbital with lower chemical potential as the vacancy $sp^2$ orbital, and the higher one as the vacancy $p_z$ orbital.
Fig.~\ref{fig:mlwfs} shows the MLWFs after this unitary rotation.

\begin{figure}
    \centering
    \includegraphics[width=0.7\textwidth]{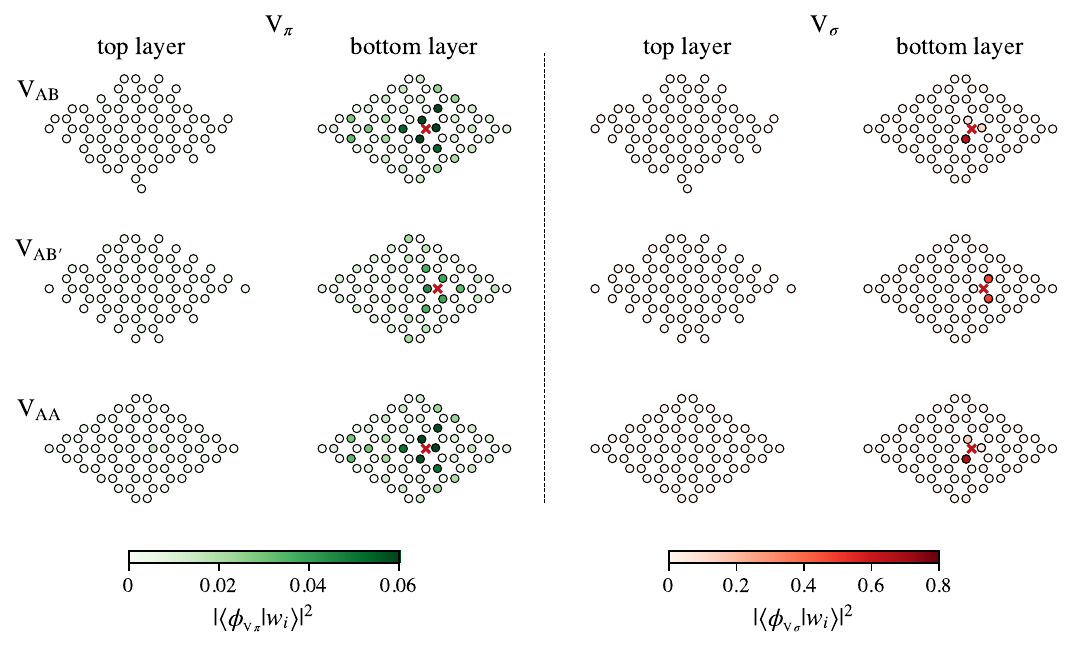}
    \caption{The probability density projected onto the MLWFs orbitals for the two vacancy states of interest, V$_\sigma$ and V$_\pi$, in V$_{\text{AA}}$ configuration.
    Here, $\ket{w_i}$ is a $(N+2)$ vector, including $(N-1)$ $\pi$ and 3 $\sigma$ MLWFs.
    The dispersion of these two vacancy states is highlighted in Fig.~\ref{fig:bands}.
    The V$_\pi$ states are quasi-localized (power-law decay) around the vacancy site, while the V$_\sigma$ states are exponentially localized at the vacancy adjacent sites. }
    \label{fig:vacancy_states}
\end{figure}

Figure~\ref{fig:vacancy_states} shows the wavefunctions of the two localized vacancy states in the spin-up channel, V$_{\pi}$ and V$_{\sigma}$, at $\Gamma$.
The V$_{\sigma}$ state is mainly localized at the lone site among the three adjacent sites.
In contrast, the V$_{\pi}$ state (``zero mode'') is quasi-localized around the vacancy site, contributed by the dominant sublattice.  
In the following analysis, we will focus on the hybridization between the V$_{\sigma}$ state and the $\pi$ bath.

\subsection{Local corrugation enhances the V$_{\sigma}$-bath hybridization} 

\begin{figure}
    \centering
    \includegraphics[width=0.9\textwidth]{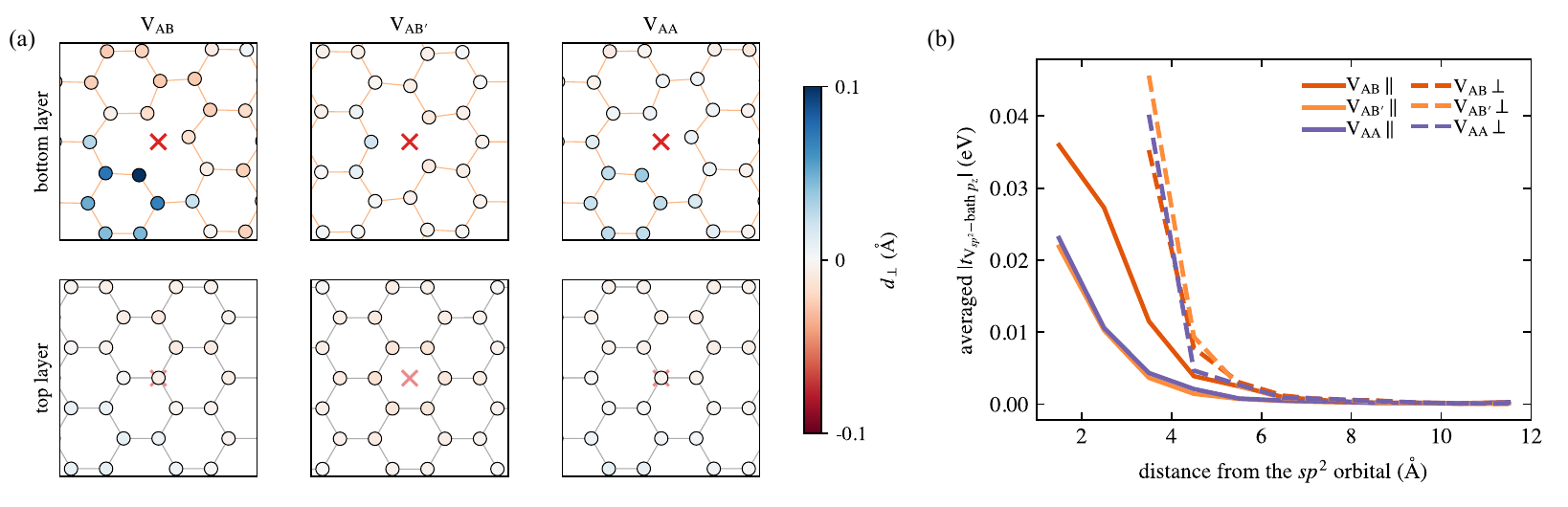}
    \caption{After the relaxation of the lattice structure, local curvature forms near the regions of the vacancy.
    This local curvature gives rise to a non-zero intra-layer hopping ($t_{\parallel}$) between the vacancy $sp^2$ orbitals and the rest of the $p_z$ orbitals. 
    (a) Position of the atoms from the bottom and top layers of the three configurations, with the atoms colored by their $z$ coordinate away from the averaged position. 
    (b) The absolute values of the intra- ($t_{\parallel}$) and inter-layer hoppings ($t_{\perp}$) between the three vacancy $sp^2$ orbitals and the $p_z$ orbitals, versus the distance away from the vacancy $sp^2$ orbitals. 
    The values were averaged over the three vacancy-adjacent sites and coarse-grained with respect to the distance. 
    We see that the intra-layer and inter-layer hoppings are of a similar order of magnitudes at equilibrium configurations.
    }
    \label{fig:local_curvature}
\end{figure}

We have also studied how local corrugation gives rise to non-zero hybridization between the V$_\sigma$ states and the $\pi$ bath in bilayer graphene.
Figure~\ref{fig:local_curvature}(a) shows the atomic positions near the vacancy for the three equilibrium configurations.
Compared to the other two cases, V$_{\text{AB}^\prime}$ causes more local curvature on the top layer rather than the bottom layer where the vacancy resides.
Figure~\ref{fig:local_curvature}(b) shows a distribution of the intra- and inter-layer $sp^2$-$p_z$ hoppings versus the distance away from the vacancy $sp^2$ orbitals.
All the absolute values of the hoppings were collected and then averaged over a given bin at some distance.
For V$_{\text{AA}}$ and V$_{\text{AB}^\prime}$ configurations, the nearest intra-layer hoppings are about 0.02~eV, and nearest inter-layer hoppings are halved, about 0.01~eV.
For V$_{\text{AB}}$, the intra-layer hoppings are slightly larger, and the nearest inter-layer hoppings are 0.035~eV.
This result is consistent with the estimated hybridization strength $V$ for each configuration, as listed in Table~\ref{table:formation_energy_spin_moment}, i.e., the hybridization between the V$_\sigma$ and the $\pi$ bath is mainly contributed by intralayer hoppings, which are caused by the local corrugation near the vacancy in the layer where the vacancy resides.
The fact that the equilibrium structure is not completely flat allows the $sp^2$ and $p_z$ orbitals to hybridize. 
Particularly, from the last two rows of the table, we see that the V$_\sigma$-$\pi$ bath hybridization strength is further enhanced with the presence of the second layer that breaks the local mirror symmetry of the bilayer, in addition to the small local curvature induced by the vacancy. 
However, the finite hybridization strength gives a low Kondo temperature, which is not attainable in experiments.
Therefore, similar to the situation in the single layer~\cite{jiang_inducing_2018}, the local corrugation introduced by the substrate can help greatly increase the Kondo temperature. 
See Section~\ref{section:local_corrugation} for a detailed study of the effect of local corrugation near the vacancy.

\begin{figure}
    \centering
    \includegraphics[width=0.95\textwidth]{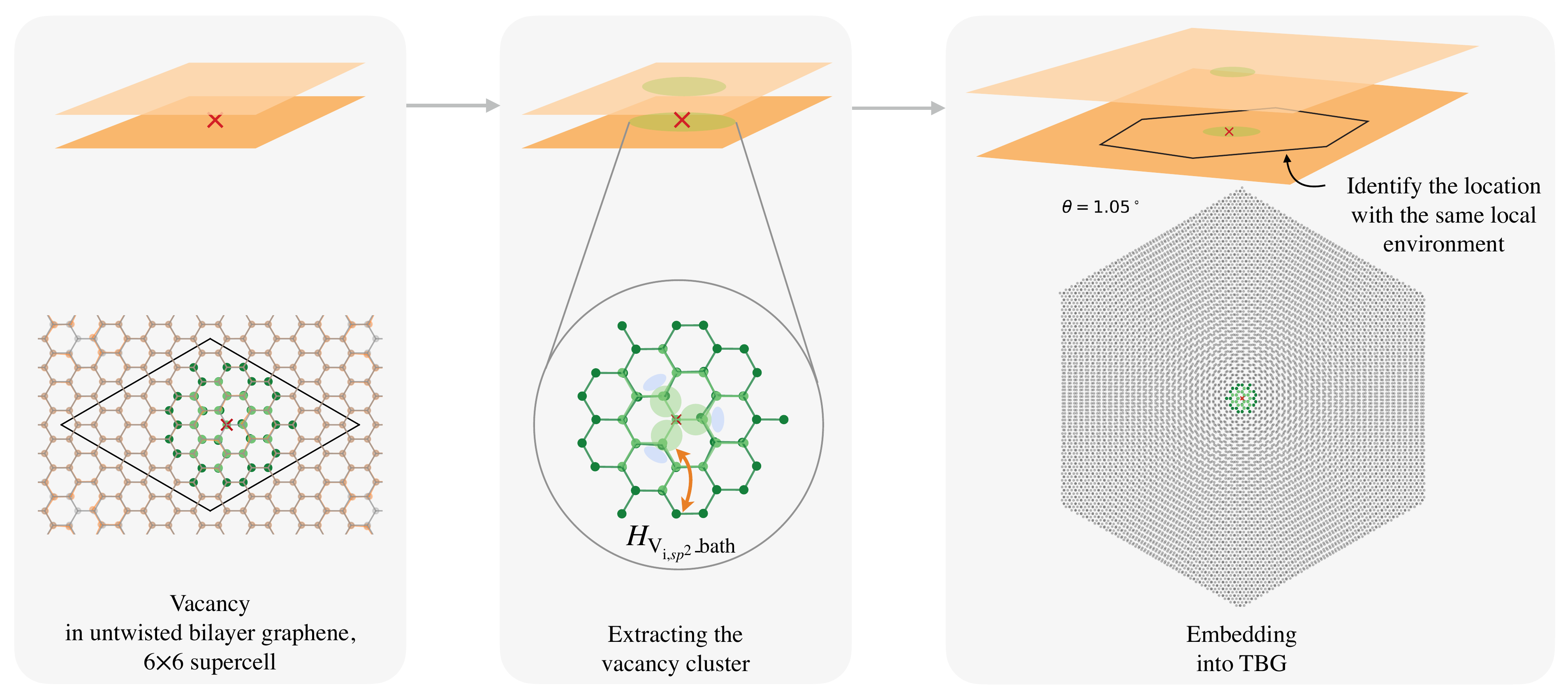}
    \caption{A schematic of the workflow of modeling dilute impurities in moir\'e heterostructures, along with the example of applying it to modeling vacancies in TBG (in this work).
    1. We start from simulating a certain type of impurity or vacancy in an untwisted bilayer structure (e.g., monovacancy in the AA-stacked bilayer graphene).
    The dark and light green sites represent the positions of the bath $\pi$ orbitals from the bottom and top layers, respectively.
    2. We then downfold to a model of all the $p_z$ orbitals on all the sites, and the $sp^2$ orbitals on the three adjacent sites of the vacancy. 
    We identify the cluster of atoms near the vacancy, both on the bottom layer and on the top layer, since they undergo significant in-plane lattice reconstruction and are mostly affected by the presence of the vacancy. 
    3. We then identify the corresponding location in a TBG that has the same local environment as in the bilayer graphene, either AA- or AB-stacked, and substitute all the hopping parameters near the vacancy with the ones obtained from the Wannierization. 
    }
    \label{fig:workflow}
\end{figure}

% ab, ab2, aa ordering
\begin{center}
\begin{table}
\renewcommand{\arraystretch}{1.5}
\begin{tabular}{m{4.5cm} m{1.4cm} m{1.4cm} m{1.3cm} | m{1.3cm} m{1.3cm} m{1.3cm} |m{1.3cm} m{1.3cm} m{1.3cm}}
 Twist angle & \multicolumn{3}{c}{$1.05^\circ$} & \multicolumn{3}{c}{$1.35^\circ$} & \multicolumn{3}{c}{$2.0^\circ$}\\
  &  V$_{\text{AB}}$ & V$_{\mathrm{AB^\prime}}$ & V$_{\text{AA}}$ & V$_{\text{AB}}$ & V$_{\mathrm{AB^\prime}}$ & V$_{\text{AA}}$ & V$_{\text{AB}}$ & V$_{\mathrm{AB^\prime}}$ & V$_{\text{AA}}$\\[0.1cm]
  \hline
Hybridization strength $V$\,(eV)  &&&&&&&&&\\
in [-200,\,200]\,meV window 
& 0.211(45) & 0.135(21) & 0.101(12) 
& 0.192(65) & 0.069(13) & 0.092(12) 
& 0.166(43) & 0.075(9) & 0.071(12) \\ 

in [-20,\,20]\,meV window 
& 0.152(16)  & 0.094(15)  & 0.106(2) 
& 0.131(13) & 0.062(4)  & 0.081(4) 
& 0.157(6) & 0.075(3) & 0.070(2)\\
\end{tabular}
\label{table:hyb_for_tbg}
\caption{Computed hybridization strength $V$ between the V$_{\sigma}$ and the $\pi$ bath of TBG at various twist angles for the three vacancy configurations at equilibrium, within two chosen energy windows.}
\end{table}
\end{center}

\subsection{General framework of the multi-scale modeling of dilute impurities in moir\'e heterostructures}
\label{section:workflow}

In our work, we have developed a general framework for modeling dilute impurities in moir\'e heterostructures. 
In the following section, we explain the details of this method with dilute vacancies in TBG as an example, which is the system of interest in our work.
However, this method is readily generalizable for building the Anderson impurity model for impurities in any other moir\'e heterostructures.
In TBG with commensurate twist angles, moir\'e unit cells usually include hundreds or thousands of atoms, making it computationally demanding for direct ab-initio simulations. 
Our method combines a much simpler ab-initio simulation of the impurity states on a smaller length scale and a description of the pristine moir\'e bath, circumventing the process of ab-initio downfolding the entire system from scratch.

Figure~\ref{fig:workflow} shows the schematic workflow of this procedure.
We start from a DFT calculation of a fully relaxed system of one vacancy in a $6\times 6$ supercell of AA-stacked bilayer graphene (see supplementary section~\ref{suppsec:DFT_monovacancy_in_bl}).
We then Wannierize the bands near the Fermi level by downfolding to the $p_z$ orbitals of the bath, and the $sp^2$ orbitals of the three adjacent vacancy sites.
We then extract the hopping terms between the vacancy $sp^2$ orbitals, V$_{i,sp^2}$ ($i=1,2,3$) and the rest of the $\pi$ orbitals at the $(N-1)$ bath sites.
We note that the hoppings among the sites that are close to the vacancy site (``vacancy cluster'') are mostly affected by the presence of the vacancy.
The wavefunction $\ket{\phi_{\text{V}\sigma}}$ of the vacancy state of interest, V$_{\sigma}$, is obtained by diagonalizing the 3$\,\times\,$3 block Hamiltonian of the V$_{i,sp^2}$ basis.
Then, we diagonalize the atomic-scale tight-binding model $H_{\text{TBG}}$ for the pristine TBG as given by Ref.~\cite{pathak_accurate_2022} at twist angle $\theta$ and obtain the eigenstates $\ket{\psi_{n\mathbf{k}}}$, written in the $\pi$ orbital basis.

In order to compute the hybridization function $\Delta(\omega)$, we will need to first compute the transition matrix elements $V_{n\mathbf{k}}$ between the vacancy states and mode $\ket{\psi_{n\mathbf{k}}}$.

Finally, we identify the location of the vacancy in TBG by searching for a local environment that is almost the same as that in the untwisted bilayer graphene.
This method could be generalized to vacancies located in intermediate stacking regions.
Here, this is easily achieved by identifying the AB or AA twist centers in TBG, then finding the atomic sites in TBG that have the closest local environment with each ``bath" site in the untwisted bilayer graphene.
We then replace all the original hoppings in the vacancy cluster in TBG with the ones we have derived from Wannierization, chosen to be within 6~{\AA} away from the vacancy site at the bottom layer, and 4\,{\AA} away at the top layer.
As a result, we constructed the hopping terms between V$_{i,sp^2}$ and the $\pi$ orbitals in TBG, i.e., $H_{\text{V}\sigma\text{-bath}}$.
Therefore, by projecting the TBG eigenstates only onto the $(N-1)$ bath $\pi$ orbitals, the tunneling elements between $\ket{\phi_{\mathrm{V}\sigma}}$ and the $n$th TBG eigenstates at $\mathbf{k}$ is given by 
\begin{equation}
V_{n\mathbf{k}} = \braket{\phi_{\text{V}\sigma} | H_{\text{V$\sigma$-bath}} \mathcal{P}_{\text{bath-TBG}} |\psi_{n\mathbf{k}}}.    
\end{equation}
The hybridization function between the V$_{\sigma}$ state $\ket{\phi_{\text{V}\sigma}}$ and the TBG $\pi$ bath is given by
\begin{equation}
    \Delta_{\text{micro}}(\omega) = \pi\sum_{n, \mathbf{k}}|V_{n\mathbf{k}}|^2 
    \delta ( \omega - E_{n\mathbf{k}} ) , 
\end{equation}
where the subscript ``$\text{micro}$" denotes that this hybridization function is obtained from the atomic-scale, microscopic model of the TBG combined with ab-initio downfolded hopping terms obtained from DFT calculations.

We estimated the hybridization strengths $V$ between the V$_\sigma$ state and the $\pi$ bath in TBG, within a chosen energy window around the Fermi level, using
\begin{equation}
    \Delta_{\text{micro}}(\omega) \approx \pi V^2 
    \sum_{n,\mathbf{k}}|\braket{\phi_{\text{V}_{\sigma}} | \psi_{n\mathbf{k}}}|^2\delta(\omega-E_{n\mathbf{k}}),
\end{equation}
which is equivalent to equation~3 in the main text.
The averaged $V$ is listed in Table~\ref{table:hyb_for_tbg}, where the numbers in the brakets represent the standard deviations of the $V(\omega)$ within the chosen energy window.
They are of similar values with the hybridization strengths in untwisted bilayer graphene.
Curiously, we noticed that both V$_{\text{AB}}$ and V$_{\text{AB}'}$'s $V$ values show a larger standard deviation, compared with V$_{\text{AA}}$, especially as the twist angle approaches the magic angle $1.05^\circ$.
We think this is because of the multifractal nature of the wavefunction in magic-angle TBG, which leads to very discontinuous flat-band wavefunctions in the atomic scale. 
Therefore, for these two cases, a simple frequency-independent $V$ may not be sufficient to describe the hybridization function $\Delta(\omega)$.
(Therefore, further NRG calculations with a more accurate $\Delta(\omega)$ need to be performed, which is beyond the scope of this study.)

Note that we used the same embedded cluster (see the green atoms in Fig.~\ref{fig:workflow}) for all these calculations.
Therefore, the effect of the local curvature change in the vicinity of the vacancy due to different twist angles was not taken into account. 
We expect this approximation to apply to small twist angles where, for example, the V$_{\text{AA}}$ vacancy only sees its local environment as AA.
The AA and AB regions get small in TBG with large twist angles. 
Therefore, in this case, the local environment of a vacancy located at an AA (or AB) center can not be well approximated by AA- (or AB-) stacked untwisted bilayer graphene.

\subsection{Local corrugation away from equilibirum}
\label{section:local_corrugation}

\begin{figure}
    \centering
    \includegraphics[width=\textwidth]{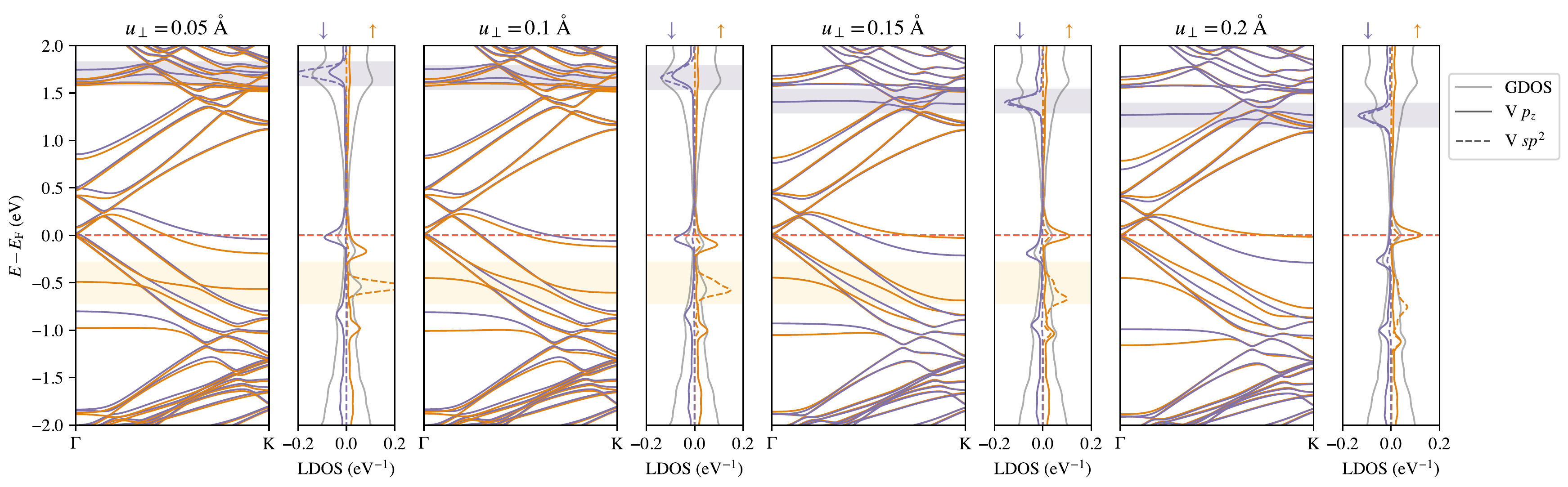}
    \caption{The band structures of the untwisted bilayer graphene with V$_{\text{AA}}$ type vacancy  with local vertical displacements from 0.05\,{\AA} to 0.2\,{\AA} (see Fig.~1f for how the three vacancy-adjacent atoms are displaced).
    The highest occupied (lowest unoccupied) V$_\sigma$ state is identified for $u_{\perp}\,<\,0.2\,${\AA} and highlighted in yellow (purple).
    With $u_{\perp}=0.2\,${\AA}, a localized V$_{\sigma}$ state is not identifiable due to the strong hybridization with the $\pi$ bath.}
    \label{fig:bands_w_curv}
\end{figure}

\begin{figure}
    \centering
    \includegraphics[width=\textwidth]{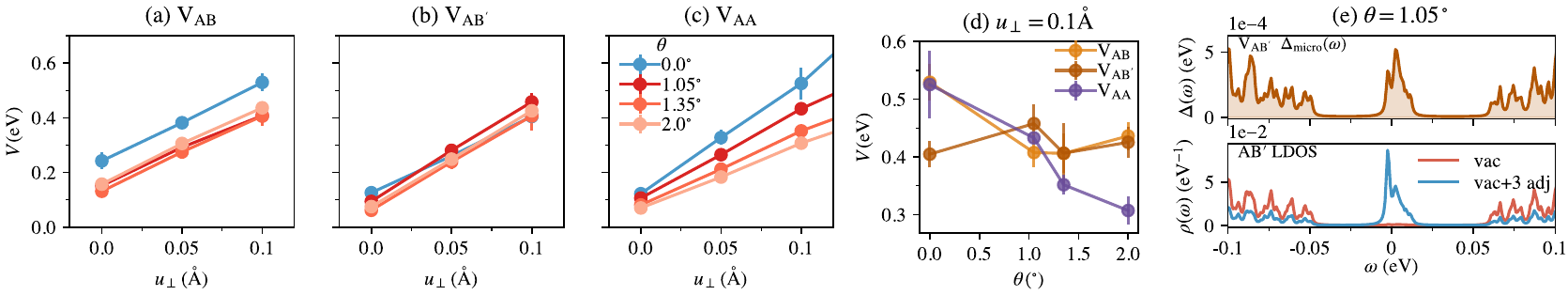}
    \caption{(a-c) The computed hybridization strength for the three vacancy types, with ad-hoc local corrugations.
    (d) $V$ versus the twist angle at $u_{\perp}=0.1\,$\AA. 
    The hybridization strength of V$_{\mathrm{AA}}$ vacancy is more sensitive to the twist compared to the other two AB-type vacancies.
    (e) The hybridization function for the $V_{\mathrm{AB}'}$ in equilibrium configuration (upper panel) and the LDOS at the corresponding vacancy site, and the LDOS sum for the vacancy site and the three adjacent sites. We can see that the vacancy site of V$_{\mathrm{AB}'}$ has almost zero contributions to the flat band DOS; when computing the value $V$, it would have a large uncertainty if one were to divide $\Delta(\omega)$ by the LDOS at only the vacancy site. Instead, we use the averaged LDOS of the vacancy and the three adjacent sites.}
    \label{fig:v_vs_curv}
\end{figure}

In a free-standing (no substrate) magic-angle TBG, the two layers naturally corrugate to lower its total energy, such that the interlayer distance in the AA regions is larger, and that in the AB regions is smaller.
This natural corrugation is shown in Fig.~2(c) in the main text, where the positions of the C atoms in the TBG at $\theta=1.05^\circ$ given by MD calculations are plotted, color-coded by their interlayer distances. 
However, we do not expect this natural corrugation introduced by twisting to enhance the hybridization significantly. 
This vertical corrugation is $\lesssim$\,$0.1$\,\AA~\cite{krongchon_registry-dependent_2023} in the moir\'e scale, and is therefore $<$\,$0.01$\,\AA  ~near the vacancy. 
In order to achieve an experimentally detectable Kondo temperature ($>$\,$1$\,K) for the vacancies in the AA regions, one needs to introduce local corrugations, especially near the vacancy, which greatly enhances the hybridization strength between the vacancy state V$_{\sigma}$ and the bath.
This could be achieved via coupling to a substrate that interacts slightly stronger with TBG compared with the widely used hBN~\cite{jiang_inducing_2018}. 

To systematically study the effect of local corrugation in enhancing the V$_{\sigma}$-bath hybridization $V$ in untwisted bi-layer graphene, and how this property gets preserved in TBG, we manually introduced vertical displacement for the three adjacent atoms near the vacancy and performed similar calculations as shown in section~\ref{section:workflow}.
We chose multiple vertical displacements $u_{\perp}$, then extracted $V$ for the untwisted bilayer graphene, and TBG with chosen twist angles.
Specifically, we moved the isolated, unbonded C atom upper by $u_{\perp}$ and the other two bonded atoms lower by $u_{\perp}$.
This fashion of introducing local corrugation is ad hoc, which is merely an attempt to mimic the substrate-induced local corrugation that was observed in experiments~\cite{jiang_inducing_2018} and to maximize the hybridization between the V$_\sigma$ states and the $\pi$ bath.
To justify this, we performed two calculations: A. only displacing the isolated C atom by $+u_{\perp}$, and B. displacing the isolated C atom by $+u_{\perp}$ and the other two atoms by $-u_{\perp}$. 
For the V$_{\mathrm{AA}}$ vacancy with $u_{\perp}=0.5$\,\AA, configuration A gives a hybridization strength 1.25(2)\,eV, smaller by $\sim 24\%$ than that given by configuration B.
Therefore, we chose to displace the atoms using configuration B in order to maximize the hybridization strength.

The $V$ values for V$_{\text{AA}}$ were presented in Fig.~1(g).
Here, we show the band structure plots for the chosen corrugations in Fig.~\ref{fig:bands_w_curv}. 
We see directly from the energies of the V$_{\sigma}$ states in the spin up and down channels that the value of Hubbard $U$ slightly decreases by $\sim 0.1\,$eV as $u_{\perp}$ increases to 0.1\,{\AA} (note the position of the lowest unoccupied spin-down V$_\sigma$ state as highlighted in the light purple box), and the onsite energy $\epsilon$ almost does not change.
However, the value of the total magnetic moment $\mu_{\mathrm{tot}}$ decreases from 1.31\,$\mu_{\mathrm{B}}$ to 1.17\,$\mu_{\mathrm{B}}$ at $u_{\perp}=0.1\,$\AA.
In Fig.~\ref{fig:v_vs_curv}, we summarized the hybridization strengths for the three vacancy types as a function of the vertical displacement and the twist angle. 
Specifically, for V$_{\mathrm{AB}'}$, the value of $V$ is computed as $V = \sqrt{\Delta(\omega)/\bar{\rho}(\omega)}$, where $\bar{\rho}(\omega) = \frac{1}{4}\sum_{i\in\lbrace \text{vac, 3~adj-sites}\rbrace} \rho_{i}(\omega)$, instead of just the LDOS at the vacancy site, which otherwise gives a huge uncertainty in $V$ since the LDOS at the vacancy site almost vanishes near $E_{\mathrm{F}}$ (see Fig.~\ref{fig:v_vs_curv}(e)).

\subsection{Comparison between the microscopic model and the lattice model}

\begin{figure}
    \centering
    \includegraphics[width=0.9\textwidth]{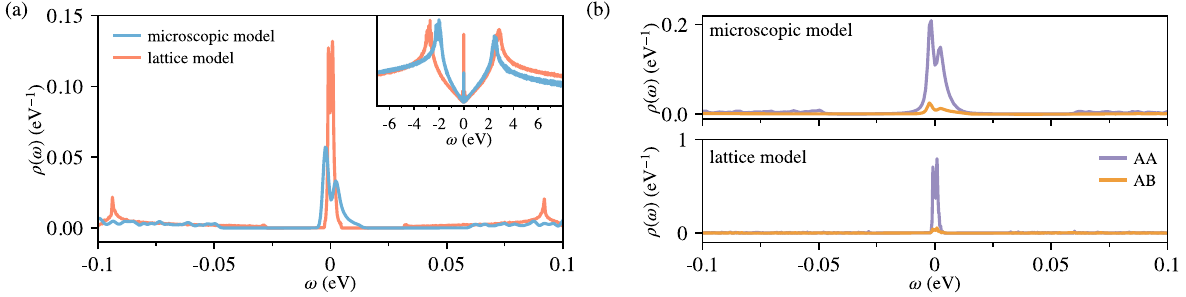}
    \caption{A comparison of the GDOS and LDOS in AA and AB sites given by the two models: $\theta=1.05^\circ$ microscopic model and $w=0.11$\,eV ($\alpha/\alpha_{\text{M}}=1.0$) lattice model.
    Both these two models agree on the overall features of the GDOS (panel (a)) and LDOS (panel (b)).
    In NRG calculations, we chose to use the GDOS and LDOS produced by the lattice model to construct the hybridization function, which allows us to reach a larger system size and achieve higher resolution near the Fermi level.}
    \label{fig:models_compare}
\end{figure}

For NRG calculations, we used the parameters computed from our microscopic modeling of the TBG+vacancy system and the local density of states given by the lattice model (described in section~\ref{app:TBGlattice}) to construct the hybridization function.
The lattice model allows for finer resolution of the hybridization function near the Fermi level since a larger system size is more attainable.
Figure~\ref{fig:models_compare} shows a direct comparison between the GDOS and the LDOS of the two models ($\theta=1.05^\circ$ microscopic model and $w=0.11$\,eV lattice model).
Both these two models agree on the overall features of the GDOS, i.e., they reproduce the flat band of a small bandwidth (20\,meV for the microscopic model, 8\,meV for the lattice model).
As for the LDOS, both models reproduce the correct features in the AA and AB sites that the flat bands are more localized at the AA sites than the AB sites.

\section{Lattice model of TBG}\label{app:TBGlattice}

The impurity host is a microscopic lattice model of TBG around the charge neutrality point where the angle enters as a free parameter \cite{PhysRevResearch.2.023325}, which is derived from the Bistritzer--MacDonald continuum model \cite{bistritzer2011moire}, and it has three main components
\begin{equation}
    H_\mathrm{host}=H_0+H_{T0}+H_{T1}.
\end{equation}
%\fk{Should we say that we consider the host at charge neutrality?}
Here, $H_0$ denotes the lattice model of two independent Dirac cones, representing the two independent layers of graphene
\begin{equation}
H_0=\sum_{\mathbf{r},l}t\left[c^{\dagger}_{\mathbf{r},l}\sigma_x c_{\mathbf{r},l}+\sum^2_{j=1}(c^{\dagger}_{\mathbf{r+a_j},j}\sigma^{+}c_{\mathbf{r},l}+\mathrm{H.c.})\right],
\end{equation}
where $t=2.8\,\mathrm{eV}$, $\mathbf{r}$ labels points on the triangular lattice, $c_{\mathbf{r},l}=(c_{\mathbf{r},\text{A},l},c_{\mathbf{r},\text{B},l})^T$ a pseudospinor operator, and layer $l=1,2$ with sublattices $\text{A},\text{B}$. The tunneling between layers in real space is separated into two parts 
\begin{equation}
\begin{split}
    H_{T_0}&=\sum_{\mathbf{r}}[c^{\dagger}_{\mathbf{r},2}\mathcal{T}_0(\mathbf{r})c_{\mathbf{r},1}+\mathrm{H.c.}],\\
    H_{T_1}&=\sum_{\mathbf{r}}\sum^6_{n=1}\left[(-1)^n c^{\dagger}_{\mathbf{r+a_n},2}\mathcal{T}_1\left(\mathbf{r}+\frac{\mathbf{a}_n}{2}\right)c_{\mathbf{r},1}+\mathrm{H.c.}\right],
\end{split}
\end{equation}
where $H_{T_0}$ denotes the interlayer tunneling at site $\mathbf{r}$, while $H_{T_1}$ represents tunneling to the nearest neighbors on the triangular lattice of the other layer ($\mathbf{a}_n, n=1,2,\ldots,6$ are nearest-neighbor lattice vectors, $\mathbf{a}_1=(\sqrt{3}/2,3/2)d$, $\mathbf{a}_2=(-\sqrt{3}/2,3/2)d$, $\mathbf{a}_3=(-\sqrt{3},0)d$, and $\mathbf{a}_j=-\mathbf{a}_{j-3}$ for $j=4,5,6$. $d$ is the nearest C atom distance. The lattice constant is $a_0=|\mathbf{a}_j|=\sqrt{3}d$. The tunneling matrices are given by
\begin{align}
\mathcal{T}_0(\mathbf{r}) &= \sum^3_{j=1}
\begin{pmatrix}
w_0\cos(\xi_{j,-}) & w_1\cos(\zeta_{j,-})\\
w_1\cos(\zeta_{j,+}) &  w_0\cos(\xi_{j,+})
\end{pmatrix}
, \nonumber \\
\mathcal{T}_1(\mathbf{r})&= \frac{1}{3\sqrt{3}}\sum^3_{j=1}
\begin{pmatrix}
w_0\sin(\xi_{j,-}) & w_1\sin(\zeta_{j,-})\\
w_1\sin(\zeta_{j,+})) &  w_0\sin(\xi_{j,+})
\end{pmatrix},
\end{align}
where $w_0, w_1$ denotes the AA and AB tunneling, respectively. In the following, we define $w=w_1$ and fix the relation $w_0=0.75w$. The abbreviations are
\begin{align}
\xi_{j,\pm} & = \mathbf{q}_j\cdot\mathbf{r}+\phi_j \pm \tfrac{1}{2}\theta
, \nonumber \\
\zeta_{j,\pm} & = \mathbf{q}_j\cdot\mathbf{r}\pm\tfrac{2}{3}\pi(j-1) +\phi_j,
\end{align}
where $\theta$ is the twist angle contained in the twist wavevector $k_\theta=2k_D\sin(\theta/2)$ with $k_D=4\pi/(3\sqrt{3}d)=4\pi/(3a_0)$.
%, which we fix at $\theta_M=1.05^{\circ}$. 
The three vectors are controlled by the $k_\theta$ as $q_1 = k_{\theta}(0,-1)$, $q_2 = k_{\theta}(\sqrt{3}/2,1/2)$ and $q_3 = k_{\theta}(-\sqrt{3}/2,1/2)$.  $\phi_j$ are three global random phases. To ensure the boundary tunneling strengths are consistent, we need extra constraints in the system sizes $L$ of the lattice, i.e., $\mathrm{mod}(q_j(\theta) L_{x|y}, 2\pi)\approx 0$ for each $\theta$. At the fixed angle $\theta=1.05^\circ$, we choose $L\equiv L_x=L_y=569a_0$. The boundary condition is taken to be a twisted boundary condition, i.e., a random twisted phase $\psi$ on the hopping matrix elements at the (right) boundaries $t\to t e^{i\psi}$.

Note that the particle-hole symmetry cannot be preserved in lattice models of TBG \cite{PhysRevB.103.205412} from the topology point of view. For each value of $w$, we shift the chemical potential (upon averaging twisted boundary conditions) to ensure charge neutrality. 

\section{Kernel polynomial method}
We computed the density of states using the kernel polynomial method (KPM). 
It is an approximation that expands a function using Chebyshev polynomials,
\begin{equation}
    f(x) = \frac{1}{\pi\sqrt{1-x^2}}\left[g_0\mu_0 + 2\sum_{n=1}^{\infty}g_n\mu_n T_n(x) \right],
\end{equation}
where $T_n(x)$s are the Chebyshev functions and $T_n(x) = \textrm{cos}(n \textrm{arccos}(x))$, $\mu_n = \int^1_{-1} f(x)T_n(x)$ are the KPM expansion moments with Jackson kernel $g_n$ \cite{Weisse2006}.
For the case of the global density of states (GDOS), the variable is $E$ and we have,
\begin{equation}
    \mu_n = \int^1_{-1} \rho(E)T_n(E) = \frac{1}{D}\sum_{k=0}^{D-1} \bra{k}T_n(H)\ket{k}=\frac{1}{D}\mathrm{Tr}(T_n(H)),
\end{equation}
where $E$ is normalized to $[-1,1]$, and the trace is evaluated stochastically with the number of random vectors $N_R$ for the $D$ states. 
The local density of states (LDOS) at site $i$ is given by
\begin{equation}
    \mu_n =  \frac{1}{D}\bra{i}T_n(H)\ket{i}.
\end{equation}
To compute the density of states (DOS) with lattice size $569$\,$\times$\,$569$, we used 2 random vectors ($N_R=2$), and a number of expansions $N_C = 2^{18}$.
The results were then averaged over a set of twisted boundary conditions, each of which was generated as a two-dimensional random phase vector and multiplied to the boundary Hamiltonian matrix entree.

The Wilson parameters for the NRG with certain logarithmic discretization $\Lambda$ can be obtained from the GDOS/LDOS once the moments $\mu_n$ is computed \cite{wu_aubry_2022}. 
The key quantities are the integration of zeroth and the first moment of energies over logarithmic bins 
\begin{equation}
    \alpha_m^{\pm} = \pm \! \int_{\pm\tilde{\epsilon}_{m+1}}^{\pm\tilde{\epsilon}_{m}} \!\! \tilde{\Delta}(\tilde{\epsilon}) \, d\tilde{\epsilon} , \quad
    \beta_m^{\pm} = \pm \! \int_{\pm\tilde{\epsilon}_{m+1}}^{\pm\tilde{\epsilon}_{m}} \!\! \tilde{\epsilon} \, \tilde{\Delta}(\tilde{\epsilon}) \, d\tilde{\epsilon},
\end{equation}
where the tilde denotes the normalized energy and hybridization (GDOS/LDOS) and the logarithmic discretization $\tilde{\epsilon}_{m+1} < \pm\tilde{\epsilon} < \tilde{\epsilon}_m$, where
\begin{equation}
\label{eq:teps_m}
  \tilde{\epsilon}_0 = 1, \qquad
  \tilde{\epsilon}_m = \Lambda^{1-z-m} \quad \text{for } m = 1, \, 2, \, \ldots.
\end{equation}
Using $T_n(x) = \cos(n\arccos x)$, and defining $\theta_m = \arccos\tilde{\epsilon}_m$ with $0\le \theta_m \le \pi/2$ for $m = 0,\, 1, \, 2, \, \ldots$, one can show that
\begin{equation}
\label{eq:alpha_m-KPM}
    \alpha_m^{\pm} = \frac{1}{\pi} \biggl[ g_0 \mu_0 (\theta_{m+1}-\theta_m) + 2 \sum_{n=1}^{N_C-1} \frac{(\pm 1)^n}{n} g_n \mu_n ( \sin n\theta_{m+1}-\sin n\theta_m ) \bigg],
\end{equation}
and
\begin{equation}
\label{eq:beta_m-KPM}
    \begin{split}      
    \beta_m^{\pm} & = \frac{1}{\pi} \biggl[ g_1 \mu_1 ( \theta_{m+1} - \theta_m ) + \sum_{n=1}^{N_C-2} \frac{(\pm 1)^n}{n} ( g_{n-1} \mu_{n-1} + g_{n+1} \mu_{n+1} ) \, ( \sin n\theta_{m+1} - \sin n\theta_m ) \\
    & \qquad \quad + \; \sum_{n=N_C-1}^{N_C} \frac{(\pm 1)^n}{n} g_{n-1} \mu_{n-1} ( \sin n\theta_{m+1} - \sin n\theta_m ) \biggr].
    \end{split}
\end{equation}      
Equations \eqref{eq:alpha_m-KPM} and \eqref{eq:beta_m-KPM} are then inserted into the standard NRG unitary transformation to yield the Wilson-chain coefficients $\varepsilon_n$ and $t_n$ \cite{Bulla2008}.

%%%%%%%%%%%%%%%%%%%%%%%%%%%%%%%%%%%%%%%%%%%%%%%%%%%%%%%%%%%%%%%%%%%%%%%%%%%

\begin{figure}
    \centering
    \includegraphics[width = 0.99\textwidth]{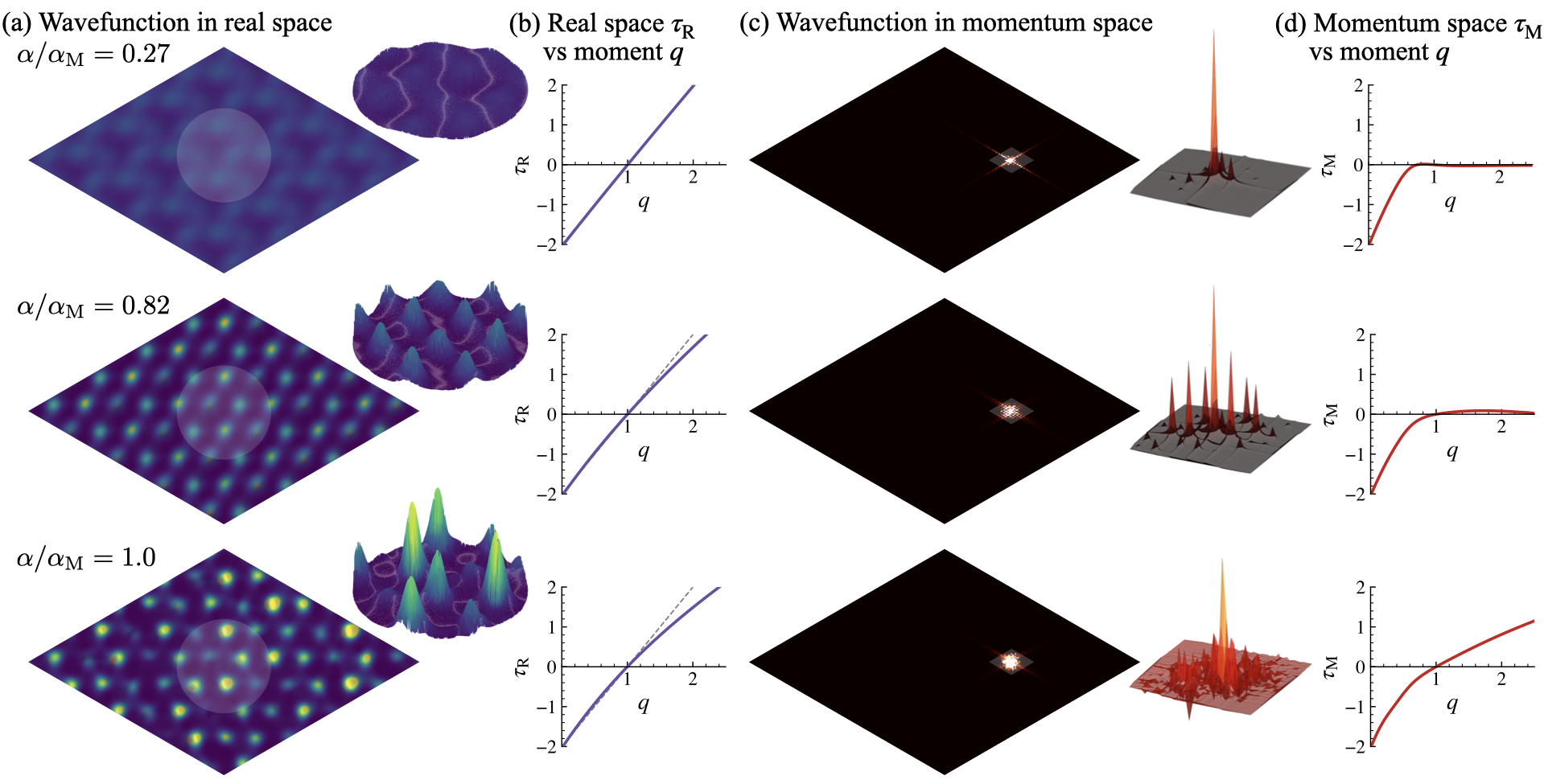}
     \caption{Multifractal nature of the TBG wavefunction at the magic angle. 
    (a) The wavefunctions in real space for $\alpha/\alpha_{\text{M}}$ at
0.27, 0.82, and 1.0, along with the zoom-in to a few moir\'e cells.
    (b) The scaling exponent $\tau_{\text{R}}$ versus moment $q$, computed using the real-space $q$th moment of the wavefunction with binning sizes $B_1=8$, $B_2=16$.
    The dashed lines represent $\tau_{\text{R}} = 2(q-1)$ (the plane wave limit).
    (c) The Fourier transform of the wavefunctions to
the momentum space, along with the zoom-in to the moir\'e Brillouin zone. 
    (d) The scaling exponent $\tau_{\text{M}}$ versus moment $q$, computed using
the momentum-space $q$th moment of the wavefunction with binning sizes $B_1=4$, $B_2=6$.}
    \label{fig:lattice_model_wf}
\end{figure}

\subsection{Wavefunction from the lattice model of TBG}
\label{section:wf_of_tbg}

In TBG, as the twist angle is tuned across the magic angle, the wavefunction also changes to a critical multifractal state, which is neither expanded nor localized~\cite{kettemann_kondo_2012, fu2020magic}. 
In Fig.~\ref{fig:lattice_model_wf}(a) and (c), we show the wavefunction probability density $|\Psi(\mathbf{r})|^2$ (only at the bottom layer), in the real space and in the momentum space, for TBG described using the lattice model, with $\alpha/\alpha_{\text{M}}=0.27,0.82,1.0$.
When the system is away from the magic angle (top panel, 0.27), the wavefunction is very delocalized and, therefore, shows a large peak at the moir\'e $\bar{K}$ point. 
As the system approaches the magic angle ($\alpha/\alpha_{\text{M}} = 1.0$), the wavefunction becomes more localized and shows more satellite peaks around the original peak at $\bar{K}$.

The multifractal nature of the wavefunction is characterized by the $q$th moment of the wavefunction probability density in either the real space or the momentum space. 
\begin{equation}
    \mathcal{I}_{\text{R}}(q) = L^d \sum_{\mathrm{r}} |\Psi(\mathbf{r})|^{2q},~~
    \mathcal{I}_{\text{M}}(q) = \sum_{\mathbf{k}}|\Psi(\mathbf{k})|^{2q}, 
\end{equation}
where $\Psi(\mathbf{k}) = \frac{1}{\sqrt{N}}\sum_{\mathbf{r}} e^{i\mathbf{k}\cdot \mathbf{r}} \Psi(\mathbf{r})$, and $\mathbf{k}$ is the k grid commensurate with the real-space lattice sites (no periodic boundary conditions are assumed for the incommensurate lattice model, though).  
Here, $L$ is the system's linear scale, and $d$ is the spatial dimension (2 in the case of TBG).
When $q=2$, this definition is consistent with the common definition of the inverse participation ratio.

For a plane wave state, with the normalization of the wavefunction taken into account, one can see that $\mathcal{I}_{\text{R}}(q) \sim L^d \times L^{qd} = L^{d(1-q)}$.
Let us define the scaling exponent $\tau_{\text{R}}$ as $\mathcal{I}_{\text{R}}(q) \sim L^{-\tau_{\text{R}}}$, we can see that for a plane wave state, $\tau_{\text{R}}(q) = d(q-1)$~\cite{kettemann_kondo_2012}.
In contrast to the constant $d$ value, the critical multifractal state is characterized by that $\tau_{\text{R}}(q) = d_{q}(q-1)$, where $d_{q}$ depends on the power $q$ of the moments. 

Figure~\ref{fig:lattice_model_wf}(b) and (d) show the scaling exponents computed using real-space wavefunction and momentum-space wavefunction, $\tau_{\text{R}}$ and $\tau_{\text{M}}$, as a function of $q$.
For $\alpha/\alpha_{\text{M}}=1.0$, $\tau_{R}$ deviates from the linear relationship of a plane-wave-like wavefunction, and $\tau_{\text{M}}$ also shows a strong dependence on $q$.
For the other two cases away from the magic angle, $\tau_{\text{R}}$ is closer to the plane wave $2(q-1)$ line (denoted by the dashed lines in panel (b)), and $\tau_{\text{M}}$ freezes for $q>2$, showing the ballistic character of the wavefunction in momentum space (i.e., the wavefunction is quite spread out in real space).
The $\tau_{\text{R/M}}$ values were calculated using the method described below (see also the supplemental note 5 of Ref.~\cite{fu2020magic}).

To compute the scaling exponent $\tau_{\text{R/M}}$, we introduce a binning factor $B$ and recompute the wavefunction averaged in the real-space bins (or on a coarse k grid).
To extract $\tau$, the following approximation is used for two consecutive binning factors $B_1$ and $B_2$, which yields similar results with fitting $\ln\mathcal{I}_{\text{R/M}}(q, B)$ vs $\ln B$ to a line, 
\begin{equation}
    \tau_{\text{R/M}}(q; B_1, B_2) = \frac{\ln\mathcal{I}_{\text{R/M}}(q, B_1) - \ln\mathcal{I}_{\text{R/M}}(q, B_2)}{\ln B_1 - \ln B_2}. 
\end{equation}

The implications of the multifractal nature of TBG wavefunctions are profound, and the physics can only be accessible through the incommensurate lattice model, but not the commensurate tight-binding model.
It has been shown in several disordered models~\cite{dobrosavljevic_kondo_1992, Miranda_kondo_1996, cornaglia_universal_2006, kettemann_critical_2009, kettemann_kondo_2012, miranda_disorder-mediated_2014, gammag_distribution_2016, slevin_multifractality_2019}, and some quasiperiodic settings~\cite{andrade_non-fermi-liquid_2015, wu_aubry_2022}, at Anderson localization transitions, the multifractal wavefunctions lead to physical observables acquiring broad distributions.
This arises because multifractal wavefunctions have probability amplitudes that are spatially distributed in a highly non-trivial fashion. 
In our work, the exact distribution of the Kondo temperatures is obtained using NRG, and also shows a long tail into the lower temperatures.
Using our method, we have access to all the possible locations in either the AA or the AB regions of TBG that highly mimic the realistic scenario, it is potentially possible to verify our predictions using STM experiments combined with vacancies created by in-situ Helium sputtering on TBG~\cite{coe2024cryogenfree}.

%%%%%%%%%%%%%%%%%%%%%%%%%%%%%%%%%%%%%%%%%%%%%%%%%%%%%%%%%%%%%%%%%%%%%%%%%%%

\begin{figure}
    \centering
    \includegraphics[width=0.48\textwidth]{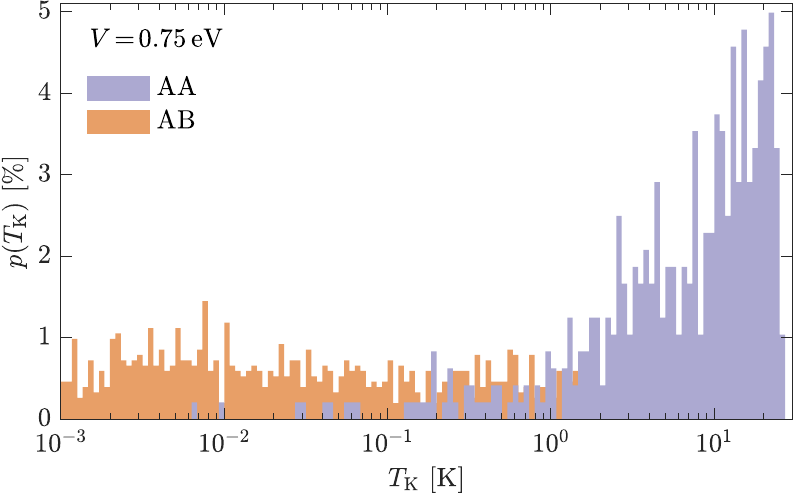}
    ~
    \includegraphics[width=0.48\textwidth]{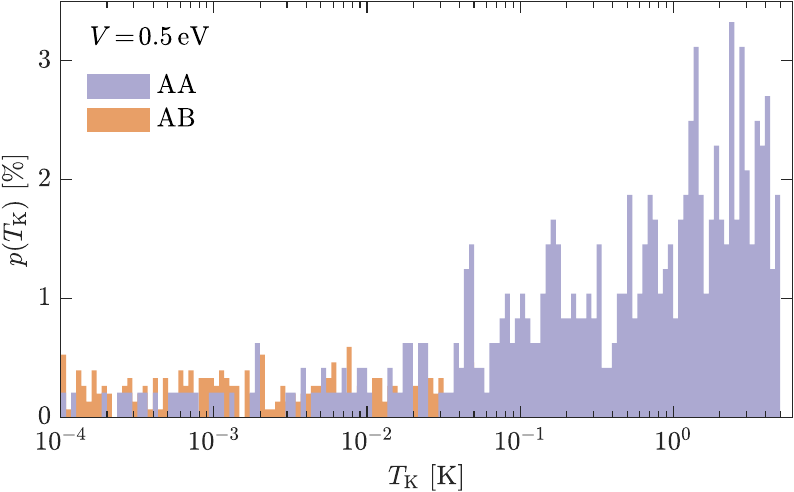}
    \caption{Distribution of $T_{\mathrm{K}}$ for different impurity locations across magic-angle TBG, analogously to Fig.~4 in the main text, but at smaller hybridization strengths $V \!=\! 0.75\,\mathrm{eV}$ and $V \!=\! 0.5\,\mathrm{eV}$.}
    \label{fig:TK_lower_V}
\end{figure}

\begin{figure}
    \centering
    \includegraphics[width=0.97\textwidth]{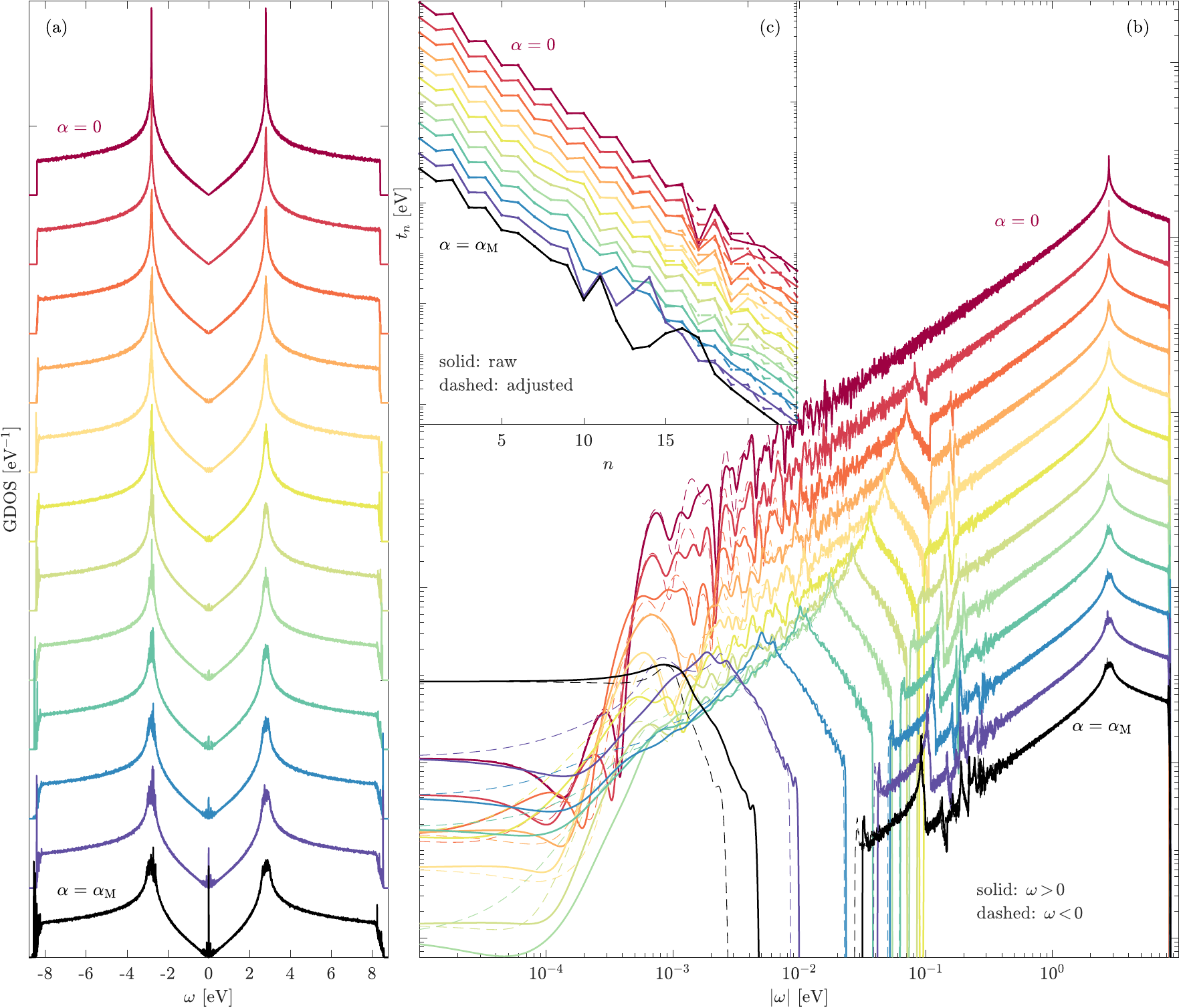}
    \caption{GDOS for various twist angles on (a) linear and (b) logarithmic scales. (c) Corresponding hopping amplitudes along the Wilson chain. All lines are shifted vertically for visualization.}
    \label{fig:GDOS_tn_KPM_NRG}
\end{figure}

\section{Numerical renormalization group}

We used an NRG discretization parameter of $\Lambda=3$ and kept up to $2000$ SU(2) multiplets during the iterative diagonalization. 
The impurity contribution to the spin susceptibility was computed as a thermodynamic property (we average over several values of $\bar{\beta}$ on the order of 1, cf.\ Eq.~(45) in Ref.~\cite{Bulla2008}). 
The local spin susceptibility was computed within the full density-matrix NRG \cite{Peters2006,Weichselbaum2007}.
To demonstrate how sensitive $T_{\mathrm{K}}$ is to $V$, we plot in Fig.~\ref{fig:TK_lower_V} the distribution of Kondo temperatures, analogously to Fig.~4 in the main text, but at two lower values of $V$.

Figure~\ref{fig:GDOS_tn_KPM_NRG} shows the GDOS for various twist angles from $\alpha \!=\! 0$ to $\alpha_{\mathrm{M}}$ on linear and logarithmic scales, together with the corresponding hopping amplitudes along the Wilson chain $t_n$.
All lines are shifted vertically for visualization. 
Consider $\alpha < \alpha_{\mathrm{M}}$ first.
From the log-log plot of the GDOS, one sees that all curves behave as $\rho(\omega) \propto |\omega|$ for low frequencies, but the KPM resolution is low below $|\omega| \lesssim 10^{-2}\, \mathrm{eV}$ and breaks down for $|\omega| \lesssim 10^{-3}\, \mathrm{eV}$. Accordingly, the Wilson chain hopping parameters $t_n$ become irregular for $n \gtrsim 15$ and spuriously flat for $n \gtrsim 20$ (as the GDOS levels off below the resolution limit). For $\alpha$ closer $\alpha_{\mathrm{M}}$, more low-energy states are available to KPM and the low-energy resolution becomes better.
At $\alpha_{\mathrm{M}}$, $\rho$ is actually flat at low energy and the $t_n$ actually follow the asymptotic $\Lambda^{-n/2}$ behavior, so the resolution limit is not seen. To overcome the KPM resolution limit for $\alpha < \alpha_{\mathrm{M}}$ and access arbitrarily small energy scales (for all $\alpha$), we adjust and extend the Wilson chain parameters using the analytically known asymptotic behavior as in Eqs.~(28) and (29) of Ref.~\cite{Bulla1997} for $\alpha \!<\! \alpha_{\mathrm{M}}$ and Eq.~(32) of Ref.~\cite{Bulla2008} for $\alpha \!=\! \alpha_{\mathrm{M}}$.
Note that this is equivalent to extrapolating the hybridization function to very low energies. 

\bibliography{refs}